\documentclass{jfm}
\usepackage{amsmath}
\usepackage{graphicx}
\usepackage{verbatim}
\usepackage{bm}

\begin{document}
\shorttitle{Methodology to characterize the energy transfer and inter-scale transport}
\shortauthor{D. Foti}
\title{A methodology to characterize the energy transfer and inter-scale transport of coherent structures using mode decomposition}

\author{Daniel Foti
 \corresp{\email{dvfoti@memphis.edu}}}
\affiliation{
  Department of Mechanical Engineering, University of Memphis, Memphis, TN 38152, USA           
  }
\maketitle
\begin{abstract}
A methodology is developed to quantify the transfer and transport of kinetic energy of specific scales associated with coherent structures.  Coherent motions are characterized by the triple decomposition of a multi-scale flow and used to define mean, coherent, and random kinetic energy.  Specific scales of individual coherent structures are identified through mode decomposition, whereby the total coherent velocity is separated into a set of velocities classified by the scale of the mode based on frequency to embed spectral characteristics.  The set of scale-specific coherent velocities are used to identify scale-specific coherent kinetic energy, which quantifies the kinetic energy of a specific scale, and an equation for the balance of scale-specific coherent kinetic energy.  Each equation includes terms with velocity triads, which represent the inter-scale transfer.   The methodology is assessed in the wake behind a square cylinder, where scales of the flow are related to the vortex shedding.  The scale-specific inter-scale transfer identifies that kinetic energy initially in the near wake is transferred from vortex shedding to scales associated with higher harmonics of the vortex shedding. In the far wake, inter-scale transfer is homogeneous and transfer occurs over many frequencies.  While the inter-scale transfer is identified over a range of scales, the scale-specific transport, convection, and dissipation of coherent kinetic energy are associated with only the largest, dominant scales.  The wake flow is subjected to unsteady inflow at various frequencies to identify the changes to inter-scale transfer and coherent kinetic energy induced by other frequencies.
\end{abstract}

\section{Introduction}\label{sec:introduction}
While turbulent flows are characterized by a broad range of scales that often appear chaotic with random fluid motions, persistent, organized coherent structures are often observed~\citep{brown1974density,hussain1985experimental}. The dynamics of coherent structures play an important role in turbulent flow behaviour~\citep{cantwell1981organized,haller2015lagrangian,hussain1983coherent}.   Various classes of coherent structures within any particular flow can exist due to different origins, disparate characteristic scales, spatio-temporal developments depending on the Reynolds number and other pertinent parameters in the flow.
Coherent structures are often the dominant flow features and appear in a wide range of flows include wall-bounded flow~\citep{jimenez2018coherent} and free shear layers~\citep{wygnanski1986large}.   However, even if they share common traits, the interactions of coherent structures among themselves or the surrounding flow as well as other associated dynamics and mechanisms such as genesis, instabilities and breakdown remain unclear.  The objective of this work is to develop a framework to elucidate and analyse the mechanisms of the interactions of coherent structures.  The interactions, herein, are defined as the transport and transfer of the kinetic energy of coherent structures in the flow.   

The transfer of kinetic energy among scales can be leveraged to identify the formation, evolution and destruction of coherent structures.  Multi-scale turbulence energy transfer has a long history based on the seminal Richardson-Kolmogorov theory~\citep{richardson1922weather,kolmogorov1941local} of the turbulent energy cascade.  The energy cascade postulates that at sufficiently large Reynolds numbers the energy transfers from the largest energy-containing scales to the smallest universal scales. The transfer of energy is described by the triadic interaction of scales, which is a result of the quadratic nonlinearity present in the Navier-Stokes equations.  The largest scales can impact on the small scale velocity statistics~\citep{mininni2006large}. However, the triadic interactions that describe the nonlinearity in the turbulence imposes significant difficulties by promoting inter-scale and nonlocal interactions~\citep{domaradzki1994energy,brasseur1994interscale} and extreme dissipation events~\citep{zhou2019extreme}.    

Energy transfer between the mean and the fluctuating parts of the flow have been studied in great detail in a variety of flows~\citep{calaf2010large,cal2010experimental,yang2015effects,gatti2018global,symon2021energy,cimarelli2016cascades}.
Energy transfer plays a key role in the organization of multi-scale coherent structures and turbulent eddies and insight into their self-sustaining mechanisms~\citep{kravchenko1993relation,hamilton1995regeneration,waleffe1997self}. 
While energy transfer between the mean and the fluctuating part (i.e., turbulence production) can provide insights for industrial scale~\citep{calaf2010large,cal2010experimental,yang2015effects,gatti2018global,symon2021energy}, the multi-scale physics lead to anisotropy~\citep{cimarelli2016cascades}, intermittency~\citep{piomelli1991subgrid,domaradzki1993analysis,cerutti1998intermittency,dubrulle2019beyond}, inhomogeneous spatial fluxes~\citep{cimarelli2016cascades}, and nonlinear redistribution of energy that is strongly scale- and position-dependent~\citep{piomelli1996subgrid,hong2012coherent}. However, quantification of turbulence production alone cannot assess these turbulence characteristics and energy transfer scenarios can be complicated by inverse cascade and energy redistribution~\citep{alexakis2018cascades,carbone2020vortex}.

Coherent structures have been shown to transfer energy to a range of scales~\citep{goto2017hierarchy,motoori2019generation}.  The inter-scale energy balances, such as the  Karman–Howarth-Monin equation and its generalizations~\cite{hill2002exact,gatti2020structure}, can be used to assess the spatial correlations in specific areas of a flow field~\citep{gomes2015energy,valente2015energy,portela2017turbulence,zhou2020energy}. However, the energy in the coherent structures cannot be isolated through traditional Reynolds decomposition, and the inter-scale interactions between coherent structures cannot be quantified.  Furthermore, the nonlinear mechanisms preclude employment of linear models to predict nonlinear energy transfer~\citep{jin2020energy,symon2021energy}. The present work develops the framework to quantify and assess the interactions of coherent structures over a range of scales. Instabilities and compact vorticity are induced in free shear layers and give rise to multi-scale coherent structures.  While the range of scales is large and pertinent at very large Reynolds numbers, in this work we focus on lower finite Reynolds numbers.  We are motivated by considering (i) that moderate and small Reynolds numbers exist in practical applications, (ii) the interactions of coherent structures with both large and small scales is likely when the inertia range is small, and (iii) the range of scales and interaction is more distinguishable at small Reynolds numbers.  

Some of the pioneering work on coherent structures was performed in free shear flows~\citep{brown1974density}. In fact, it is well-known that most free shear flows contain coherent structures, which often persist in time and space~\citep{liu1989coherent,goldstein1988nonlinear}. The time scale of the coherent structures is often much larger than the smallest scale and is dependent of the initial conditions.  The wake over a cylinder given rise to the prominent coherent structures at relatively low Reynolds number, $\mathrm{Re} = U_\infty D/\nu$, where $U_\infty$ is the upstream velocity, $D$ is the diameter, and $\nu$ is the kinematic viscosity.  A two-dimensional von K\'{a}rm\'{a}n vortex street appears in the wake at $Re \approx 45$ and becomes three-dimensional around $Re \approx 150$.  The wake becomes turbulent around $\mathrm{Re} > 10^3$.  The non-dimensional Strouhal number $\mathrm{St} = f D/U_\infty$, where $f$ is the vortex shedding frequency, increases with Reynolds number for $\mathrm{Re} < 1000$ ~\citep{luo2003characteristics,bai2018dependence}. Due to the prominence and identifiable frequency of the shedding vortices, the wake flow is well-suited to be used to identify coherent structure interactions. The nonlinear dynamics of coherent structures include spectral broadening, where interactions induce motions with slightly different frequencies~\citep{wu2012spectral}.   The coherent structures are stable and persist in far downstream.  The interactions of coherent structures and processes of energy transfer are still not fully understood in the wake flow and provide a unique test bed for understanding coherent structure interactions.

To capture the energy transfer from coherent structures, the spatio-temporal fluctuations associated with coherent structures need to be identified and separated from the total fluctuations~\citep{reynolds1972mechanics}.
A commonly employed technique to quantify the turbulent fluctuations that are associated with coherent structures is the triple decomposition of the velocity~\citep{hussain1970mechanics}.  The decomposition requires additional insights and operators to distinguish the coherent quantity.  
The triple decomposition of the velocity leads to three coupled equations that capture the evolution of kinetic energy: (i) the mean kinetic energy equation, (ii) the coherent kinetic energy equation associated only with the coherent fluctuations, and (iii) random kinetic energy equations.  The three equations have been used to identify the exchange of energy between the mean, coherent, random scales.  In particular, the coherent kinetic energy equation has been applied to analysis on scale-by-scale energy analysis~\cite{thiesset2011scale}, controls of coherent structures~\citep{chen2021theoretical}, and analysis of inter-scale transfer~\cite{reynolds1972mechanics,thiesset2014dynamical,chan2021interscale}, but they only consider the scale of one coherent motion. Further insight in needed to identify multiple, specific scales and account a scale-specific coherent kinetic energy.

While many techniques have been proposed to identify coherent structures such as Eulerian diagnostic~\citep{hunt1988eddies,jeong1995identification,dubief2000coherent}, they might not be all appropriate to quantify the coherent fluctuations.   Temporal filtering methods~\citep{hussain1986coherent} have commonly been employed to identify a single coherent structure with a regular Strouhal number, however, coarse-graining~\citep{dong2020coherent,motoori2019generation} and Lagrangian methods~\citep{haller2015lagrangian,chrisohoides2003experimental} can also be employed. However, these aforementioned techniques almost always quantify a single coherent fluctuation as part of the triple decomposition.  When multiple coherent structures with different scales are present in the flow, these techniques do not allow for analysis of the interactions of the separate coherent structures.  We pursue a methodology that allows multiple coherent structures to be represented through modal decomposition~\citep{sirovich1987turbulence,holmes2012turbulence,mezic2013analysis}.  The method leverages the large quantities of data with high spatio-temporal resolution that have become ubiquitous in flow solutions.  Mode decomposition provides a method for analysis and data reduction where the flow variable is decomposed into a tuple of amplitude, temporal coefficient, and spatial mode. When model decompositions are pair with compressive sensing~\citep{donoho2006compressed,fowler2009compressive}, the objective reductions of data can be improved~\citep{kutz2013data,jovanovic2014sparsity}.  A common mode decomposition is proper orthogonal decomposition (POD), which is the decompositions of the velocity covariance matrix and produces orthogonal modes that optimally represent the variance of the data~\citep{sirovich1987turbulence}.   Due to the orthogonality of the modes and ordered mode amplitude, this is commonly employed for analysis~\citep{berkooz1993proper,verhulst2015altering,foti2020adaptive} and reduced-order modelling~\citep{holmes2012turbulence,rowley2004model}.  However, in this work the orthogonality is undesirable because it limits the analysis of the kinetic energy distributed between two different modes.   However, dynamic mode decomposition (DMD), the approximate eigendecomposition of the operator that maps the evolution between snapshots, organizes modes based on their temporal frequency~\citep{rowley2009spectral,schmid2010dynamic}. Both analysis~\citep{sarmast2014mutual,foti2018similarity} and reduced-order modelling~\citep{annoni2017method,proctor2016dynamic,qatramez2021reduced} have used DMD.  There are several benefits of employing DMD to identify the interactions of coherent structures through energy transfer: (i) modes are identified by a unique frequency scale which is directly associated with coherent structure, (ii) the frequency component imparts spectral qualities while not having to be based in Fourier spectral space, (iii) modes are non-orthogonal such that the coherent kinetic energy associated with interaction of two coherent structures can be identified, and (iv) triadic interactions and energy transfer can be captured in the developed scale-specific coherent kinetic energy equation below.

In what follows, we will develop a set of equations for scale-specific coherent kinetic energy based identifying prominent DMD modes that are associated with coherent structures.  Section \ref{sec:method} formulates and described methodology of the triple decomposition with dynamic mode decomposition.  Compressive sensing is included to identify prominent modes and reduce the number of coherent interactions.  Section \ref{sec:numerical} discusses the numerical methods for computational simulation of the wake flow.  Section \ref{sec:results} details the analysis on a series of low Reynolds number wake flows that have easily identifiable coherent structures and section \ref{sec:conclusions} provides conclusions and discussion.

\section{Methodology}\label{sec:method}
The Navier-Stokes equations for an incompressible flow over a cylinder are the following ($i,j = 1,2,3$ and repeated indices imply summation):
\begin{align}
   \frac{\partial u_i}{\partial x_i} &= 0, \\
   \frac{\partial u_i}{\partial t} + u_j \frac{\partial u_i}{\partial x_j} &= -\frac{\partial p}{\partial x_i} + \frac{1}{\mathrm{Re}}\frac{\partial^2 u_i}{\partial x_j \partial x_j} + F_i,
   \label{eqn:ns}
\end{align}
where $x_i$ indicates the streamwise, vertical, and spanwise directions, $u_i$ is the velocity, $p$ is the pressure, $F_i$ are external body forces.  The Reynolds number $\mathrm{Re} = U_\infty D/\nu$ is based on relevant cylinder diameter $D$, mean inflow velocity $U_\infty$, and kinematic viscosity $\mu$.   The force term $F_i= 0$ for cases that will be developed herein.

\subsection{Transport of kinetic energy}
The total kinetic energy $E = \frac{1}{2} u_i u_i$ is obtained by multiplying the Eqn. (\ref{eqn:ns}) by $u_i$.  
The transport and transfer of kinetic energy play a key role the energy cascade and the organization of coherent structures. Analysis of the change in kinetic energy is often described in one of two forms: (i) Spectral (i.e., Fourier modes) analysis of the energy $\hat{E}(\bm{\kappa},t) = \bm{\hat{u}}\cdot\bm{\hat{u}}^*$ evolution in wavenumber space $\bm{\kappa}$ given by:
\begin{equation}
    \frac{d}{dt} \hat{E}(\bm{\kappa},t) = \kappa_l P_{jk} (\bm{\kappa}) \mathcal{R} \left ( i \sum_{\kappa^\prime} \langle \hat{\bm{u}}_j(\bm{\kappa})\hat{\bm{u}}_k^*(\bm{\kappa}^\prime) \hat{\bm{u}}_k^*(\bm{\kappa} - \bm{\kappa}^\prime)\rangle \right ) - 2\nu \bm{\kappa}^2 \hat{E}(\bm{\kappa},t)),
    \label{eqn:spectral}
\end{equation}
where bold symbols indicate vectors and the change in an energy at scale $\kappa$ is balance by the non-linear triadic transfer of energy $\hat{T}$, the first term of the RHS,~\citep{domaradzki1994energy} and dissipation $\hat{\epsilon}=2\nu \bm{\kappa}^2 \hat{E}$. The triadic interactions manifest in Fourier space as a triplet of three wavenumbers vectors ($\bm{\kappa}_1, \bm{\kappa}_2, \bm{\kappa}_3$) or frequencies ($f_1, f_2, f_3$).  This form provides a clear quantification of the kinetic energy and its transfer, including the possible reverse cascade effects, in terms of length scales.  (ii) Physical (i.e., point) analysis of the instantaneous energy transfer process of kinetic energy $E(\bm{x},t) = \frac{1}{2}\bm{u}\cdot\bm{u}$ obtained as the following in index notation where repeated indices imply summation: 
\begin{equation}
   \frac{\partial E}{\partial t} + u_i \frac{\partial E}{\partial x_i} = \frac{\partial}{\partial x_i} \left (2\nu u_j s_{ij} - \frac{u_ip}{\rho} \right) - 2 \nu s_{ij} s_{ij},
   \label{eqn:ke}
\end{equation}
where $s_{ij} = \frac{1}{2}(\partial_ju_i + \partial_i u_j)$ is the instantaneous strain rate.  This accounts for the evolution in physical space of kinetic energy due to the transport via viscous (first term on the RHS) and pressure (second term on the RHS) forces and dissipation $\epsilon=2 \nu s_{ij} s_{ij}$.  This captures the energy fluxes in space and is used to assess mean and turbulence kinetic energy evolution.   

Due to the nature of the energy evolution equations, it is difficult to quantify the transfer and transport of kinetic energy of both contributions  of a particular turbulent scale and its spatial fluxes. A common approach to separate the mean from the turbulent scale in physical space is the Reynolds decomposition of a quantity $q$: $q(\bm{x},t) = Q(\bm{x}) + q(\bm{x},t)^\prime$, where $Q$ is the mean and $q^\prime$ represents the turbulent fluctuations. Employing the Reynolds decomposition for velocity $u$ and pressure $p$, and performing an averaging procedure (designated by $\overline{\cdot}$) on Eqn. (\ref{eqn:ke}), the balance of mean kinetic energy (MKE), $K = \frac{1}{2} U_i U_i$, is obtained as the following:
\begin{equation}
    \frac{\partial K}{\partial t} + U_i \frac{\partial K}{\partial x_i} = \frac{\partial}{\partial x_i} \left ( \frac{1}{\mathrm{Re}} \frac{\partial K}{\partial x_i} - U_iP - \overline{u^{\prime}_i u^{\prime}_j}U_j \right ) + \overline{u_i^\prime u_j^{\prime}}\frac{\partial U_i}{\partial x_j} - \frac{2}{\mathrm{Re}} S_{ij}S_{ij},
    \label{eqn:mke}
\end{equation}
where the production $\mathcal{P} = -\overline{u_i^\prime u_j^{\prime}}\partial_j U_i$ is the mechanism in which energy at the mean scale is extracted to the fluctuating scale.  The mean dissipation is defined as $\mathcal{E}=\frac{2}{\mathrm{Re}} S_{ij}S_{ij}$ is a always positive and shows that the mean strain rate $S_{ij} = \frac{1}{2} (\partial_j U_i + \partial_i U_j)$ is responsible for dissipation of energy.

The balance of the fluctuating portion of the kinetic energy, the turbulence kinetic energy (TKE), $k = \frac{1}{2} \overline{u_i^\prime u_i^\prime}$, can be obtained in the usual way by subtracting the MKE from Eqn. (\ref{eqn:ke}).  The transport of TKE is the following:
\begin{equation}
    \frac{\partial k}{\partial t} + U_i \frac{\partial k}{\partial x_i} = \frac{\partial}{\partial x_i} \left ( \frac{1}{\mathrm{Re}} \frac{\partial k}{\partial x_i} - \overline{u^{\prime}_ip^{\prime}} - \frac{1}{2}\overline{u^{\prime}_j u^{\prime}_i u^{\prime}_j} \right ) - \overline{u_i^\prime u_j^{\prime}}\frac{\partial U_i}{\partial x_j} - \frac{2}{\mathrm{Re}} \overline{s_{ij}s_{ij}},
    \label{eqn:tke}
\end{equation}
The turbulence dissipation $\epsilon = \frac{2}{\mathrm{Re}} \overline{s_{ij}s_{ij}}$, is often significant in highly turbulent flows compared to the mean dissipation.  However, nonlinearity~\citep{domaradzki1994energy} in the turbulent transport term, $\frac{1}{2}\overline{u^{\prime}_i u^{\prime}_j u^{\prime}_i}$ imposes significant difficulties by promoting \emph{inter-scale and nonlocal interactions}, which is fully witnessed in terms of spectral triad in the Eqn. (\ref{eqn:spectral}).  

In order to elucidate more details about the inter-scale transfer, we will leverage the coherency in the flow based on coherent structures~\citep{hussain1986coherent}.  The effects of coherent structures can be separated from the turbulent fluctuations through triple decomposition~\citep{hussain1970mechanics} of a quantity $q$:
\begin{equation}
q(x,t) = Q(x) + \tilde{q}(x,t) + q^{\prime\prime} (x,t),
\label{eqn:coherent}
\end{equation} 
where $Q(x)$ is the average, $\tilde{q}(x,t)$ represents the coherent contribution and $q^{\prime\prime} (x,t)$ is the incoherent or random residual. The triple decomposition to the velocity and pressure are given by
\begin{align}
   u_i = U_i + \tilde{u}_i + u_i^{\prime \prime}, &\quad p = P + \tilde{p} + p^{\prime \prime}, \\
   u_i^\prime = \tilde{u}_i + u_i^{\prime \prime}, &\quad p^\prime = \tilde{p} + p^{\prime \prime}, \\
   \langle u_i \rangle = U_i + \tilde{u}_i , &\quad \langle p \rangle = P + \tilde{p},
\end{align}
and follow the properties $\overline{\tilde{q}} = \overline{q^{\prime\prime}} = \tilde{q^{\prime\prime}} = 0$.  Following \citep{reynolds1972mechanics}, equations of the coherent velocity can be obtained from Eqn. (\ref{eqn:ns}) expanded with the triple decomposition by first averaging (often referred to as filtering) over the coherent scale and then subtracting the equations of the mean velocity.  The equations of the coherent velocity are the following:
\begin{align}
    \nonumber\frac{\partial \tilde{u}_i}{\partial t} + U_j \frac{\partial \tilde{u}_i}{\partial x_j} + \tilde{u}_j\frac{\partial U_i}{\partial x_j} = - \frac{\partial \tilde{p}}{\partial x_i} + \frac{1}{\mathrm{Re}} \frac{\partial^2 \tilde{u}_i}{\partial x_j \partial x_j} + \frac{\partial}{\partial x_j} \left ( \overline{\tilde{u}_i\tilde{u}_j} - \tilde{u}_i \tilde{u}_j \right ) \\ 
    - \frac{\partial}{\partial x_j} \left ( \widetilde{u^{\prime\prime}_i u^{\prime\prime}_j} - \overline{u^{\prime\prime}_i u^{\prime\prime}_j} \right).
    \label{eqn:coh_u}
\end{align}

Similarly, the equations for the random velocity are obtained by removing Eqn. (\ref{eqn:coh_u}) from Eqn. (\ref{eqn:ns}) to obtain the following:
\begin{align}
    \nonumber\frac{\partial u^{\prime\prime}_i}{\partial t} + U_j \frac{\partial u^{\prime\prime}_i}{\partial x_j} + \tilde{u}_j \frac{\partial u^{\prime\prime}_i}{\partial x_j} + u^{\prime\prime}_j\frac{\partial U_i}{\partial x_j} &+ u^{\prime\prime}_j\frac{\partial \tilde{u}_i}{\partial x_j} = \\
    &- \frac{\partial p^{\prime\prime}}{\partial x_i} + \frac{1}{\mathrm{Re}} \frac{\partial^2 u^{\prime\prime}_i}{\partial x_j \partial x_j} + \frac{\partial}{\partial x_j} \left ( \overline{u^{\prime\prime}_i u^{\prime\prime}_j} - u^{\prime\prime}_i u^{\prime\prime}_j \right ).
    \label{eqn:rand_u}
\end{align}

Despite that both Eqns. (\ref{eqn:coh_u}) and (\ref{eqn:rand_u}) are not closed, we still gain considerable insights from the energy transferred between the different scales.  The triple decomposition of the average kinetic energy is given as the sum of the average of each component as follows:
\begin{equation}
   E = \frac{1}{2}\overline{U_i U_i} + \frac{1}{2}\overline{\tilde{u}_i\tilde{u_i}} + \frac{1}{2}\overline{u^{\prime\prime}_i u^{\prime\prime}_i}.
\end{equation}

The equations for the evolution for all three components of the average kinetic energy can be obtained by multiplying the corresponding momentum equations by $U_i$, $\tilde{u}_i$ and $u^{\prime\prime}_i$ and averaging.   
The transport of MKE in Eqn. (\ref{eqn:mke}) remains unchanged despite the triple decomposition.  However, turbulent production is split between production the coherent scales $\mathcal{P}_c$ and production to the random scales $\mathcal{P}_r$, demonstrating that energy is transferred to both coherent and random scale.  The separation of the turbulent production is show as the following:  
\begin{equation}
    \overline{u_i^\prime u_j^\prime}\frac{\partial U_i}{\partial x_j} = \mathcal{P}_c + \mathcal{P}_r = \overline{\tilde{u}_i \tilde{u}_j}\frac{\partial U_i}{\partial x_j} + \overline{u_i^{\prime\prime} u_j^{\prime\prime}}\frac{\partial U_i}{\partial x_j}.
\end{equation}

The coherent kinetic energy is given as 
\begin{equation}
    \overline{\tilde{k}} = \frac{1}{2} \overline{\tilde{u}_i\tilde{u}_i}
\end{equation}
contains the total energy that is present in the coherent motions in the flow.  
The evolution of coherent kinetic energy (CKE) can be obtained from algebraic manipulation of $\tilde{u}_i$ multiplied by Eqn. (\ref{eqn:coh_u}) and averaging as the following:
\begin{align}
    \nonumber\frac{ \partial \overline{\tilde{k}}}{\partial t} + U_i \frac{\partial \overline{\tilde{k}}}{\partial x_i} = \frac{\partial}{\partial x_i} \left ( \frac{1}{\mathrm{Re}} \frac{\partial \overline{\tilde{k}}} {\partial x_i} - \overline{\tilde{u}_i\tilde{p}} - \frac{1}{2}\overline{\tilde{u}_j \tilde{u}_i \tilde{u}_j} - \overline{\widetilde{u^{\prime\prime}_i u^{\prime\prime}_j} \tilde{u}_j} \right ) \\ -  \overline{\tilde{u}_i \tilde{u}_j}\frac{\partial U_i}{\partial x_j} + \overline{\widetilde{u^{\prime\prime}_i u^{\prime\prime}_j}\frac{\partial \tilde{u}_i}{\partial x_j}} - \frac{2}{\mathrm{Re}} \overline{\tilde{s}_{ij}\tilde{s}_{ij}}.
    \label{eqn:cke}
\end{align}
We will identify each term in Eqn. (\ref{eqn:cke}) as
\begin{equation}
   \mathcal{A}_t + \mathcal{A} = \mathcal{T}_v - \mathcal{T}_p - \mathcal{T}_t - \mathcal{T}_r - \mathcal{P}_c + \mathcal{P}_{cr} - \tilde{\epsilon},
   \label{eqn:short_cke}
\end{equation}
where several additional terms appear compared to Eqn. (\ref{eqn:tke}).  These include $\mathcal{T}_r$, the transport due to random fluctuations, and $\mathcal{P}_{cr}$, production of random kinetic energy from the coherent strain rate. The latter represents the transfer of energy from the coherent scales to the random scales.  

The random kinetic energy (RKE) is computed by the random velocity as
\begin{equation}
    \overline{k^{\prime\prime}} = \frac{1}{2} \overline{u^{\prime\prime}_i u^{\prime\prime}_i},
\end{equation}
and the balance of RKE obtained from Eqn. (\ref{eqn:rand_u}) is the following:
\begin{align}
    \nonumber\frac{\partial k^{\prime\prime}}{\partial t} + 
    U_i \frac{\partial k^{\prime\prime}}{\partial x_i} + 
    \overline{\tilde{u}_i \frac{\partial \widetilde{k^{\prime\prime}}}{\partial x_i}}= 
    \frac{\partial}{\partial x_i} \left ( \frac{1}{\mathrm{Re}} \frac{\partial k^{\prime\prime}}{\partial x_i} - \overline{u^{\prime\prime}_i p^{\prime\prime}} - \frac{1}{2}\overline{u^{\prime\prime}_i u^{\prime\prime}_j u^{\prime\prime}_i}  \right ) \\ -  
    \overline{u^{\prime\prime}_i u^{\prime\prime}_j}\frac{\partial U_i}{\partial x_j} -
    \overline{\widetilde{u^{\prime\prime}_i u^{\prime\prime}_j}\frac{\partial \tilde{u}_i}{\partial x_j}} - 
    \frac{2}{\mathrm{Re}} \overline{s^{\prime\prime}_{ij}s^{\prime\prime}_{ij}},
    \label{eqn:rke}
\end{align}
where production from both the mean and coherent scales are present.

\subsection{Identification of coherent structure scales}
In order to separate the coherent quantity from the random quantity, coherent structure identification techniques are require beyond what is needed to separate the mean and the fluctuating quantity.  Often the method used to isolate the coherent quantity is based on a single scale and employ interval or phase averaging based on a specific frequency.  
However, turbulent flows are characterized by a broad range of scales and disparately-sized coherent structures, each with different time scales. Some methods such as fluctuation analysis~\cite{chrisohoides2003experimental,foti2016wake} and mode decomposition~\cite{berkooz1993proper,schmid2010dynamic} can reveal many scales.  Mode decompositions, specifically those related to Koopman theory~\cite{mezic2013analysis} provide a mathematical basis to identify scale related to many coherent structures based on their spectral characteristics. 

Koopman operator theory is a formalism that allows us to relate the observations on a system to the underlying state-space dynamics.  In particular, we seek a Koopman-invariant subspace on the space of $\bm{u}$ that can be used to reduce the nonlinear spatio-temporal dynamics to a linear combination of time-evolving spatial modes through a Koopman mode decomposition.  The Koopman operator $K_\tau$ acts on the observable of the state space of the flow $\bm{g}(\bm{u})$ to map the time evolution such that 
\begin{equation}
    K_\tau \bm{g}(\bm{u}(\bm{x},t)) = \bm{g}(\bm{u}(\bm{x},t+\tau)).  
\end{equation}
The Koopman operator is a linear operator, which allows us to analyze its eigendecomposition and spectrum. The Koopman eigenfunction $\psi^l_i(\bm{u})$ are identified by a eigenvalue, $\lambda^l$, which we associate with a specific time scale. The eigenfunction associated with the Koopman operator are the following:
\begin{equation}
   K_\tau \psi^l(\bm{u}(\bm{x},t)) = e^{\lambda^l t}\psi^l(\bm{u}(\bm{x},t)).
\end{equation}
The eigenvalue is associated with a specific real-valued frequency $\lambda^l = \textrm{i} \omega^l$.

The Koopman mode decomposition is enabled by the expansion of the eigenfunctions 
\begin{equation}
    K_\tau \bm{g}(\bm{u}) = \sum_{l=1}^\infty \bm{g}^l \phi^l(\bm{u})e^{\lambda^l \tau},
\end{equation}
where $\bm{g}^k$ are the Koopman modes obtained by projecting the observable into the eigenfunction.  An important aspect of Koopman mode decomposition is that if two observables are related through a linear operator, then their modes are related through the same linear operator.

In this work, we will use a numerically efficient method to calculate the mode decomposition, dynamic mode decomposition~\citep{schmid2010dynamic}.  It is a linear approximation to the Koopman operator~\citep{rowley2009spectral}, but retains the unique characteristic scales based on the frequency associated with a mode. It is designed to find the spectral characteristics of the linear operator $A$ of the discrete dynamical system $\bm{x}_{k+1} = A \bm{x}_k$, where $\bm{x}$ is vector-valued quantity.   The flow is decomposed in DMD into a tuple of scalar amplitude $\alpha^k$, complex temporal coefficient $\mu^k(t) = \textrm{i}\mu^k_i + \mu^k_r = e^{\lambda^k t} $, and spatial dynamic mode $\phi^k(\bm{x})$.   In this work, the vector-quantity observable is a instantaneous snapshot of the three components of the velocity variable and pressure variable. The algorithm employed to calculate the DMD tuple is given in Appendix \ref{sec:dmd}.

In what follows, a methodology will be developed to capture both the spectral and physical energy transport. A generalized quantity fluctuation of a coherent structure is quantified by a tuple obtained in DMD as the $l$th scale-specific quantity: 
\begin{equation}
    \tilde{q}^l = \alpha^l \phi^l \mu^l.
\end{equation}
The total contribution of $R$ scales to the coherent quantity are the sum of all scale-specific terms with associated modes as follows:
\begin{equation}  
    \tilde{q} = \sum_{l=1}^R \tilde{q}^l = \sum_{l=1}^R \alpha^l \phi^l \mu^l,
\end{equation}
and the instantaneous quantity in Eqn. (\ref{eqn:coherent}) can be written as
\begin{equation}
   q(x,t) = Q(x) + \sum_{l=1}^R \alpha^l \phi^l \mu^l + q^{\prime\prime} (x,t),
\label{eqn:coherent_mode}
\end{equation} 
where the random quantity $q^{\prime\prime}$ is calculated from the residual of the instantaneous velocity and sum of the mean and coherent velocity.  The decomposition of the scale-specific coherent velocity and pressure are as follows: 
\begin{equation}
    \tilde{u_i}^l = \phi_i^l \alpha^l \mu^l ,\quad \tilde{p}^l = \phi_p^l \alpha^l \mu^l,
\end{equation}  
and summation of $R$ modes gives the coherent velocity and coherent pressure:
\begin{equation}
   \tilde{u_i} = \sum^R_{l=0} \tilde{u_i}^l, \quad \tilde{p} = \sum^R_{l=0} \tilde{p}^l.
\end{equation}

Using the mode decomposition, the equations for the coherent velocity in Eqn. (\ref{eqn:coh_u}) can be rewritten of scale-specific quantities as follows:
\begin{align}
    \nonumber\frac{\partial \tilde{u}^l_i}{\partial t} + U_j \frac{\partial \tilde{u}^l_i}{\partial x_j} &+ \tilde{u}^l_j\frac{\partial U_i}{\partial x_j} = - \frac{\partial \tilde{p}^l}{\partial x_i} + \frac{1}{\mathrm{Re}} \frac{\partial^2 \tilde{u}^l_i}{\partial x_j \partial x_j} \\ &+ \frac{\partial}{\partial x_j}\sum_{l=0,n=0}^R \left (  \overline{\tilde{u}^l_i\tilde{u}^n_j} - \tilde{u}^l_i \tilde{u}^n_j \right )  
    - \frac{\partial}{\partial x_j} \left ( \widetilde{u^{\prime\prime}_i u^{\prime\prime}_j}^l - \overline{u^{\prime\prime}_i u^{\prime\prime}_j} \right),
    \label{eqn:coh_uk}
\end{align} 
where the second to last term incorporates the sum the correlation between all modes and the last term is projected into the space of $\tilde{u}^l$.  The mode functions are not assumed to be orthogonal so inter-scale associations are possible.  The scale-specific CKE is obtained by multiplying the $l$th and $m$th modes and averaging as follows: 
\begin{equation}
    \overline{\tilde{k}}^{l,m} = \frac{1}{2}\overline{\tilde{u}_i^l \tilde{u}_i^{m*}} = \frac{1}{2}\overline{\mu^l\mu^{m *}}\alpha^l \phi^l_i \alpha^m\phi^{m *}_i, 
    \label{eqn:ss_cke} 
\end{equation}
where $*$ is the complex conjugate and assume that $\overline{\cdot}$ averaging is temporal averaging.  The averaging assumptions could be weakened to include procedures such as ensemble averaging, in which case averaging would be performed over all three components of the mode decomposition.  However, in this work we will only focus on temporal averaging for a statistically stationary flow, $\overline{\cdot}$, only the temporal component $\mu = \mu(t)$ are subject to the averaging because $\phi_i = \phi_i(x)$ only.  
The CKE is obtained by summing over all combinations as follows:
\begin{equation}
     \overline{\tilde{k}} = \sum_{l=0}^R\sum_{m=0}^R \overline{\tilde{k}}^{l,m }.
\end{equation}
 
By multiplying Eqn. (\ref{eqn:coh_uk}) by $\tilde{u}_i^{m*}$ and time averaging, all terms in Eqn. (\ref{eqn:short_cke}) can be written as the product of modes. 
\begin{enumerate}
    \item The scale-specific temporal CKE advection captures how the interaction of two modes changes the scale-specific CKE in time.  Due to the assumption of temporal averaging of statistically stationary flows, this term is zero.  The term is defined as
    \begin{equation}
        \mathcal{A}^{l,m}_t = \frac{\partial}{\partial t} \overline{\tilde{k}}^{l,m} = \frac{1}{2}\frac{\partial}{\partial t} \left ( \overline{\mu^l\mu^{m *}}\right )\alpha^l \phi^l_i \alpha^m\phi^{m *}_i  = 0.
    \end{equation}    
    \item Scale-specific mean CKE advection describes how the mean flow advects the scale-specific CKE. The term is defined as
    \begin{equation}
        \mathcal{A}^{l,m} = U_i \frac{\partial \overline{\tilde{k}}^{l,m}}{\partial x_i} = \frac{1}{2} U_i \overline{\mu^l\mu^{m *}}\alpha^l \alpha^m \frac{\partial}{\partial x_i} \left ( \phi^l_j \phi^{m *}_j\right ).
        \label{eqn:a_ss}
    \end{equation} 
    \item Scale-specific transport of CKE via viscous forces captures the dyadic interaction of two modes associated with viscosity.  This term is 
    \begin{align}
     \mathcal{T}_v^{l,m} = \frac{1}{\mathrm{Re}} \frac{\partial^2 \overline{\tilde{k}}^{l,m}}{\partial x_i \partial x_i} 
     = \frac{1}{2\mathrm{Re}} \overline{\mu^l\mu^{m*}}\alpha^l\alpha^m  \frac{\partial^2}{\partial x_i \partial x_i} \left(\phi_j^l\phi_j^{m*} \right) .
     \label{eqn:tv_ss}
     \end{align} 
    \item Scale-specific transport of CKE by pressure is defined as
    \begin{align}
     \mathcal{T}_p^{l,m} = \frac{\partial}{\partial x_i} \left ( \overline{\tilde{u}^{m*}_i\tilde{p}^{l}} \right ) 
     = \overline{\mu^l\mu^{m*}}\alpha^l\alpha^{m} \frac{\partial}{\partial x_i} \left ( \phi_i^{m*}\phi^l_p \right ),
     \label{eqn:tp_ss}
    \end{align}
    where $\phi^l_p$ is the $l$th mode of the scalar pressure field.
    \item The scale-specific inter-scale transport via turbulence is a related to the wavenumber triad in Eqn. (\ref{eqn:spectral}) because it is the non-linear, non-local term that has contributions from three modes (or scales). The benefit of the present approach is that the equations remain in real space, but modes are directly related to scales in the flow.  The turbulent transport terms appear due to the multiplication of $u_i^{m*}$ with the second-to-last term in Eqn. (\ref{eqn:coh_uk}) before averaging as $\tilde{u}_i^{m*} \partial_j  \sum_{l=0,n=0}^R  \tilde{u}_i^{l}\tilde{u}^n_j$.  The summation captures the effects of all scales on the evolution of scale-specific CKE. After manipulation and averaging, the individual contribution from each triad is split between turbulent transport $\mathcal{T}_t^{l,m,n}$ and inter-scale transfer $\mathcal{P}_t^{l,m,n}$:
    \begin{align}
        \mathcal{T}_t^{l,m,n} + \mathcal{P}_t^{l,m,n} &= \frac{\partial}{\partial x_i}  \left (\overline{\tilde{u}^{l}_i\tilde{u}_j^{m*}\tilde{u}^n_j} \right )  - \overline{ \tilde{u}^{l}_i\tilde{u}^n_j \frac{\partial \tilde{u}_j^{m*}}{\partial x_i}} \\
        &=  \overline{\mu^l\mu^{m*}\mu^n}\alpha^l\alpha^m\alpha^n \frac{\partial}{\partial x_i} \left (\phi_i^l\phi_j^{m*}\phi_j^n \right) - \overline{\mu^l\mu^{m*}\mu^n}\alpha^l\alpha^m\alpha^n \phi_i^l \phi_j^n \frac{\partial \phi_j^{m*}}{\partial x_i},
        \label{eqn:turb_ss}
    \end{align}
    where the first term on the RHS is represents the turbulent transport and the second term on the RHS is the inter-scale transfer. The sum over all the modes in the inter-scale transfer can be reduced to $\sum_{l=0,m=0,n=0}^R \mathcal{P}_t^{l,m,n} = \frac{1}{2} \sum_{l=0,m=0,n=0}^R \mathcal{T}_t^{l,m,n}$ and the total turbulent transport term in Eqn. (\ref{eqn:cke}) is recovered.  The total effects of both the scale-specific turbulent transport and inter-scale transfer on the evolution of the scale-specific CKE is the following:
    \begin{align}
        \mathcal{T}_t^{l,m} &= \sum_{n=0}^R \mathcal{T}_t^{l,m,n}, \\
        \mathcal{P}_t^{l,m} &= \sum_{n=0}^R \mathcal{P}_t^{l,m,n}.
   \end{align}
    \item Scale-specific transport of CKE via random fluctuations captures how random velocity fluctuations affect a single coherent scale, but the correlation of the random fluctuations is projected in the space of the $l$th mode.  The term is defined as
    \begin{equation}
        \mathcal{T}_r^{l,m} = \frac{\partial}{\partial x_i} \left ( \overline{\widetilde{u^{\prime\prime}_i u^{\prime\prime}_j}^l \tilde{u}^{m*}_j} \right ) =  \frac{\partial  }{\partial x_i} \left ( \overline{\widetilde{u^{\prime\prime}_i u^{\prime\prime}_j}^l \mu^{m*}} \alpha^m \phi_j^{m*}\right ).
        \label{eqn:tr_ss}
    \end{equation}
    \item The scale-specific CKE production reveals how MKE is transferred to a specific mode (or scale) by the following:
    \begin{equation}
        \mathcal{P}_c^{l,m} = \overline{\tilde{u}^l_i \tilde{u}^{m*}_j}\frac{\partial U_i}{\partial x_j} = -\overline{\mu^l\mu^{m*}}\alpha^l \phi^l_i \alpha^m\phi^{m*}_j \frac{\partial U_i}{\partial x_j}.
        \label{eqn:prod_ss}
    \end{equation}
    \item Scale-specific RKE production from CKE identifies specific contributions of modes to produce RKE by the following follows:
    \begin{equation}
        \mathcal{P}_{cr}^{m} = \overline{u^{\prime\prime}_i \tilde{u}^{\prime\prime}_j \frac{\partial \tilde{u}^{m*}_i}{\partial x_j}} = \overline{u^{\prime\prime}_i \tilde{u}^{\prime\prime}_j \mu^{m*}} \alpha^m \frac{\partial \phi^{m*}_i}{\partial x_j}.
    \end{equation}   
    \item Scale-specific CKE dissipation is the mechanism where the interaction of two modes removes CKE from the flow by the following:
    \begin{equation}
        \tilde{\epsilon}^{l,m}_c = \frac{2}{Re} \overline{\tilde{s}^{l,m}_{ij}\tilde{s}^{l,m}_{ij}} = \frac{1}{2\mathrm{Re}} \overline{\mu^l\mu^{m*}}\alpha^l\alpha^m \left ( \frac{\partial \phi^l_i}{\partial x_j} + \frac{\partial \phi^l_j}{\partial x_i} \right )\left ( \frac{\partial \phi^{m*}_i}{\partial x_j} + \frac{\partial \phi^{m*}_j}{\partial x_i} \right ).
        \label{eqn:eps_ss}
    \end{equation}
\end{enumerate}
Overall, each term plays a role in the evolution of the scale-specific CKE and allows us to quantify both the spectral effects of coherent structures and the spatial fluxes of coherent structures.  The equation for the evolution of the scale-specific CKE is given as the following:
\begin{equation}
  \mathcal{A}^{l,m}_t + \mathcal{A}^{l,m} = \mathcal{T}^{l,m}_v - \mathcal{T}^{l,m}_p - \mathcal{T}^{l,m}_t - \mathcal{P}^{l,m}_t- \mathcal{T}^{l,m}_r - \mathcal{P}^{l,m}_c + \mathcal{P}^{m}_{cr} - \tilde{\epsilon}^{l,m}.
   \label{eqn:short_cke_ss}
\end{equation}

\section{Numerical methods}\label{sec:numerical}
\indent \indent We employ the CURVIB method~\citep{ge2007numerical} to undertake direct numerical simulations of the flow over an immersed body.  The three-dimensional, incompressible continuity and momentum equations in generalised curvilinear coordinates formulated as follows ($i,j,k,l=1,2,3$ and repeated indices imply summation):
\begin{equation}
J\frac{\partial U^{i}}{\partial \xi^{i}}=0,
\label{eqn:eq_continuity_general}
\end{equation}
\begin{align}
\frac{1}{J}\frac{\partial U^{i}}{\partial t}=& \frac{\xi _{l}^{i}}{J}\left( -%
\frac{\partial }{\partial \xi^{j}}({U^{j}u_{l}})+\frac{\mu}{\rho}%
\frac{\partial }{\partial \xi^{j}}\left(  \frac{g^{jk}}{J}\frac{%
\partial u_{l}}{\partial \xi^{k}}\right) -\frac{1}{\rho}\frac{\partial }{\partial \xi^{j}} \left(\frac{%
\xi _{l}^{j}p}{J} \right)-\frac{1}{\rho}\frac{\partial \tau _{lj}}{\partial
\xi^{j}}\right) ,
\label{eqn:eq_momentum_general}
\end{align}
where $\xi^{i}$ are the curvilinear coordinates, $\xi _{l}^{i}={\partial \xi^{i}}/{\partial x_{l}}$ are the transformation metrics, $J$ is the Jacobian of the geometric transformation, $u_{i}$ is the $i$-th component of the velocity vector in Cartesian coordinates, $U^{i}$=${(\xi _{m}^{i}/J)u_{m}}$ is the contravariant volume flux, $g^{jk}=\xi _{l}^{j}\xi _{l}^{k}$ are the components of the contravariant metric tensor, $\rho $ is the density, $\mu $ is the dynamic viscosity, and $p$ is the pressure.  
The governing equations are discretized using the three-point central, second-order accurate finite difference scheme on a hybrid staggered/non-staggered grid and integrated in time using a second-order accurate projection method employing a Newton–Krylov method to advance the momentum equation. An algebraic multigrid acceleration along with a generalized minimal residual solver is used to solve the pressure Poisson equation as described in \citet{kang2011high}.
%
%

The CURVIB method is designed to capture immersed boundaries embedded in the background domain rather than using a body-fitted grid.  The method treats boundaries on the immersed body as a sharp interface and boundary conditions are reconstructed on the grid node of the background grid.  The boundary condition is interpolated to the grid nodes in the vicinity.  Previously, the CURVIB method has been used for direct numerical simulation of cardiovasular flows~\citep{borazjani2008curvilinear} and large-eddy simulations of hydrokinetic turbines~\citep{kang2014onset} and wind turbines~\citep{foti2016wake}.  Details can be found in \citet{ge2007numerical} and \citet{kang2011high}.

A direct numerical simulations of the flow around a square cylinder with $\mathrm{Re} = U_\infty D/ \nu = 175$, where is the incoming velocity, $D$ is the width of the cylinder, and $\nu$ is the kinematic viscosity, is performed. 
At this Reynolds number the well-known von K\'{a}rm\'{a}n vortex street forms in the wake of the cylinder. 
The flow is simulated within a quasi-two-dimensional computational domain in the vertical and streamwise directions $(L_x \times L_y) = (18D \times 12D)$, with periodic boundaries in the spanwise $z$-direction.  A negligible thickness in the $L_z$ direction is included due to the three-dimensional implementation of CURVIB method. The computational domain is discretized with $(N_x \times N_y \times N_z) = (351 \times 201 \times 6)$ grid points with uniform spacing within $D$ of the square cylinder and stretching in the vertical and streamwise directions towards all of the boundaries.  Slip-wall boundary conditions are used on the upper and lower walls with an imposed incoming constant volumetric flux at the inlet boundary and a convection outflow condition. 

The simulation is integrated forward with a time step $\Delta t = 0.2$.  After an initial period where initial transients are removed, the instantaneous snapshots and flow statistics are obtained.  A total of $M = 4000$ instantaneous snapshots are collected at a uniform interval of $\Delta t$ to be used to construct the dynamic modes.  The time period of the snapshots is over 40 periodic oscillations of the von K\'{a}rm\'{a}n vortex shedding.  The simulation is run an additional $36000$ time steps, where statistics of the flow are obtained to compare convergence to the statistics over the initial 4000 snapshots used in analysis. The first, second, and third order moments of the statistics show similar convergence over the initial and total simulation duration.

\section{Results}\label{sec:results}
\begin{figure}
   \begin{center}
      \includegraphics[width=\textwidth]{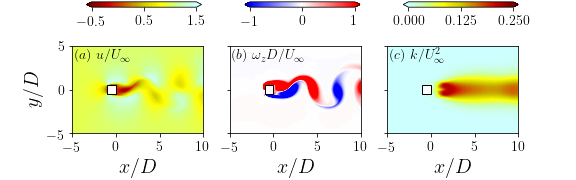}
      \caption{\label{fig:velocity} (a) Instantaneous streamwise velocity $u/U_\infty$, (b) instantaneous out-of-plane vorticity $\omega_z D/U_\infty$, and (c) turbulence kinetic energy $k/U_\infty^2$.}
    \end{center}
\end{figure}

Figure \ref{fig:velocity}(a) shows the instantaneous streamwise velocity normalized by the incoming velocity $u/U_\infty$.  The successive pairs of vortices shed from the cylinder are observed in the wake.  The out-of-plane vorticity $\omega_z D /U_\infty$ in Fig. \ref{fig:velocity}(b) captures the strength of the vortices and their alternating pattern.  While the Reynolds number of the flow is relatively low and the regime of the wake is laminar, the velocity has a fluctuating component in the wake. The fluctuating velocity is used to the quantify the turbulence kinetic energy $k = \frac{1}{2} \overline{u^\prime_i u^\prime_i}$ (TKE) averaged over 40 oscillations of the vortex shedding frequency.  Figure \ref{fig:velocity} shows that the TKE in the wake is not negligible and is about one-half of the incoming mean kinetic energy, $K_{in}/U^2_\infty = 0.5$, in the near wake. The contours of high TKE are contained primarily in the wake where the shed vortices convect. The maximum TKE appears less that a diameter behind the cylinder in the shear layer.

\begin{figure}
   \begin{center}
       \includegraphics[width=.55\textwidth]{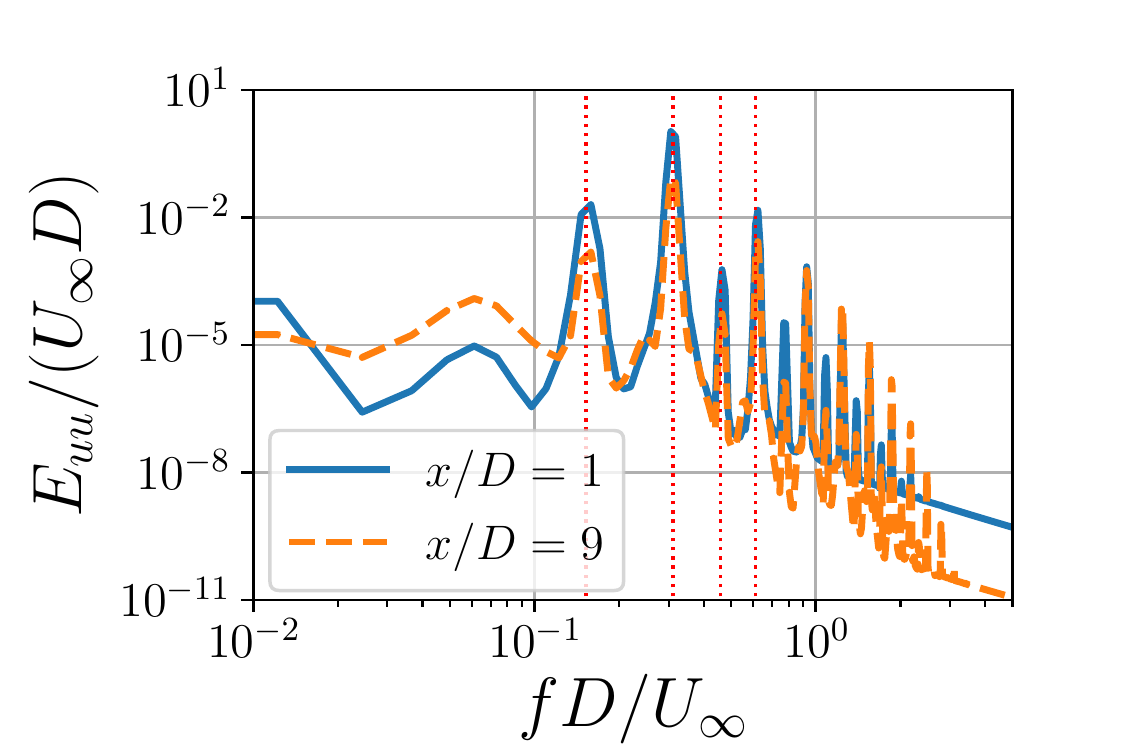}
       \caption{\label{fig:spectra} The power spectral density of the streamwise velocity $E_{uu}$ at $x/D=1$ and $x/D=9$ along the centerline.  Vertical dotted red lines indicate multiples of  $\mathrm{St}/\mathrm{St}_s = \{1, 2, 3, 4\}$.}
    \end{center}
\end{figure}

The power spectral densities of streamwise velocity component $E_{uu}$ behind the square cylinder at $x/D=1$ and $x/D=9$ along the centerline are shown in Fig. \ref{fig:spectra}.  Both spectra capture the non-dimensionalized shedding frequency $\mathrm{St}_s = f_s D/U_\infty$ of the vortices and are relatively similar to each other indicating that the spectral energy is not redistributed further downstream in the wake. The first frequency is found to be $\mathrm{St}_s=0.159$, which is similar to those identified in previous studies~\citep{sharma2004heat,sohankar1999simulation}. The shedding frequency mode and its integer multiples contain the highest energy contributions, but the wake does contain energy over the range of the resolved frequencies.  Prominent peaks include $\mathrm{St}/\mathrm{St}_s = \{1, 2, 3, 4, 5, 6\}$.  A specific frequency multiple will be identified as $i_n$, where $n$ is the integer frequency multiple, e.g., $i_1$ is the shedding frequency. Due to the nearly discrete frequency modes present in the wake, this flow provides an apt test case to identify and characterize the interactions of scales. In what follows, we will identify, classify, and elucidate details of energy transfer from the scales.

\subsection{Identification and classification of coherent structures}
\begin{figure}
   \begin{center}
       \includegraphics[width=\textwidth]{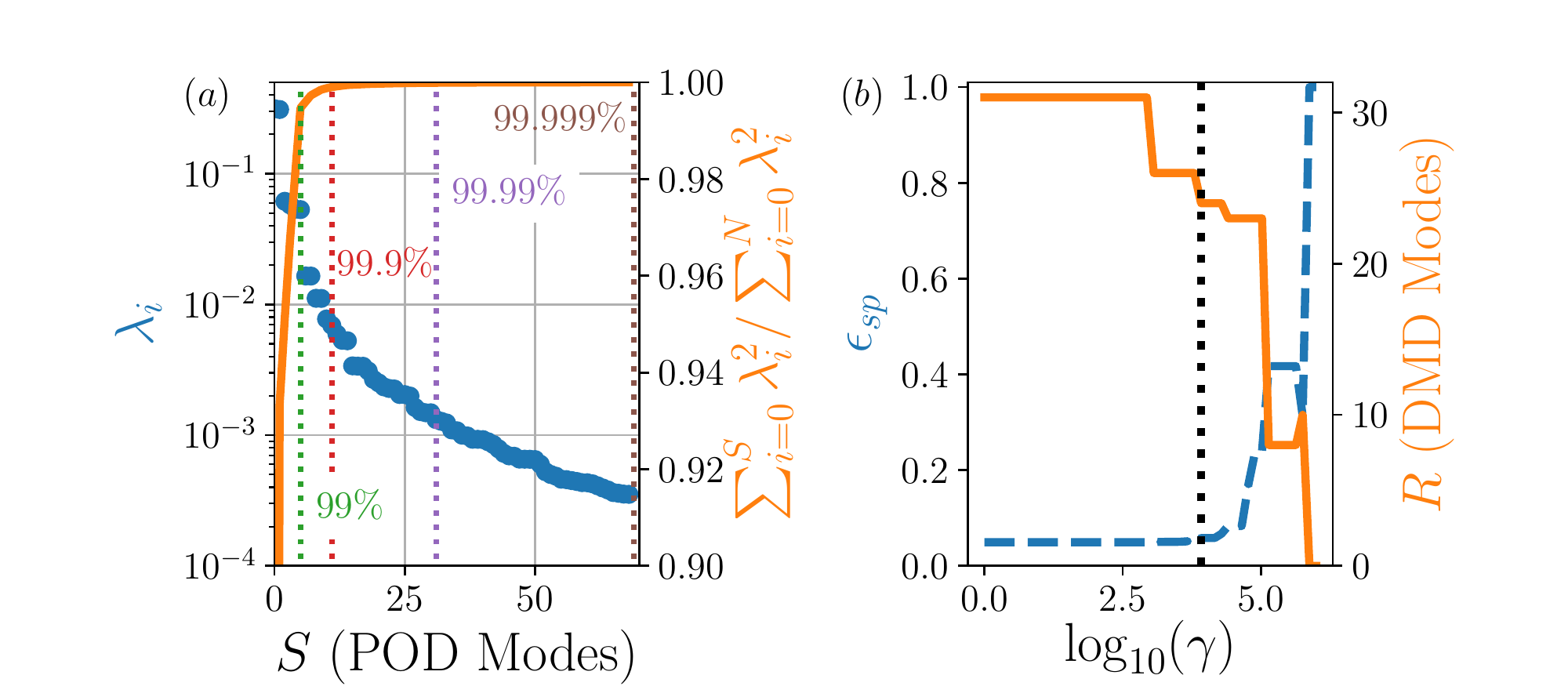}
       \caption{\label{fig:selection} (a) The singular values $\lambda$ and the cumulative energy of the POD modes and (b) the sparse sensing error $\epsilon_{sp}$ and corresponding number of DMD modes $R$ as a function of $\gamma$ for 99.99\% $(S=31)$ POD modes.}
    \end{center}
\end{figure}

The instantaneous velocity components and pressure are saved at regular intervals ($\Delta t=0.02$) over a period of 40 vortex oscillations resulting in the collection of $N=4000$ snapshots.  We perform DMD using the algorithm described in Appendix \ref{sec:dmd} with singular value decomposition (SVD) regularization. The SVD regularization, which projects the snapshot matrix into POD modes, enables dimensionality reduction to improve efficiency and remove spurious modes. Figure \ref{fig:selection}(a) shows the first $S=69$ singular values $\lambda_i$ corresponding to total cumulative energy of 99.999\%.  The first $S = 5, 11$, and $31$ POD modes correspond to 99\%, 99.9\%, and 99.99\% of the cumulative energy, respectively.   After recovering the complex frequencies $\mu^k$ and eigenvectors $y^k$, compressive sensing is employed to promote sparsity in the dynamic modes~\citep{jovanovic2014sparsity}.  Both the sparsity-promoting DMD algorithm trained on a L1 norm regularization~\citep{jovanovic2014sparsity} and a multi-task elastic net trained on a L1/L2 norm regularization~\citep{pan2020sparsity} produce similar results for sparsification of the resultant DMD modes.  The sparse sampling residual is given by 
\begin{equation}
   \epsilon_{sp} = \frac{\lVert \Sigma V^H - \Phi D_\alpha V_{and} \rVert_2} {\lVert \Sigma V^H \rVert_2},
   \label{eqn:sp_residual}
\end{equation}
where $D_\alpha = \mathrm{diag} ( \alpha )$.
Figure \ref{fig:selection}(b) shows sparse sampling residual and the corresponding number of DMD modes $R$ as a function of the sparse sampling regularization parameter $\gamma$  using the first $S=31$ POD modes, which correspond to a cumulative energy of 99.99\%.  As $\gamma$ increases, the number of modes, $R$,  selected generally reduces and the residual $\epsilon_{sp}$ increases.    

\begin{figure}
   \begin{center}
       \includegraphics[width=\textwidth]{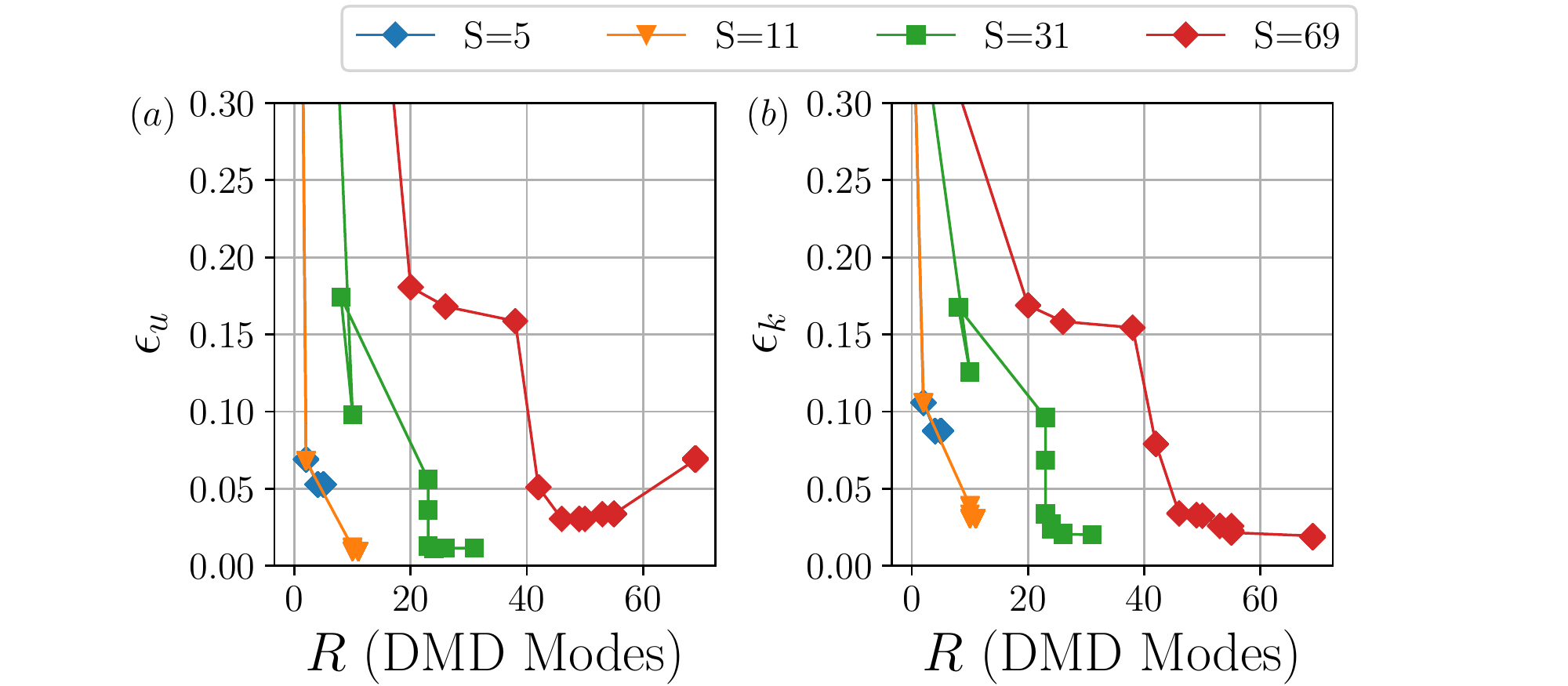}
       \caption{\label{fig:residual} DMD residuals based on (a) snapshot reconstruction $\epsilon_u$ and (b) TKE minimization $\epsilon_k$ based on the number of POD modes $S$ and number of DMD modes $R$. }
    \end{center}
\end{figure}

Further analysis is warranted for selected an appropriate degree of dimensionality reduction $S$ and compressive sensing parameter $\gamma$.  Each pair of $S$ and $\gamma$ produce different sets of DMD modes with different amplitudes because during the compressive sensing portion, the amplitudes are rescaled.  We focus on new residual metrics that are collected from a particular set of DMD modes given by a $S, \gamma$ pair.  We focus on two residual error formulations: (i.) The reconstruction L2 residual of the snapshot matrix from the $R$ DMD modes given by the following:
\begin{equation}
   \epsilon_{u} = \frac{\lVert X - \Phi D_\alpha V_{and} \rVert_2} {\lVert X \rVert_2}.
   \label{eqn:u_residual}
\end{equation}
Figure \ref{fig:residual}(a) shows the reconstruction residuals for a range of sparsity-promoting DMD solutions projected into different POD modes: $S=5, 11, 31,$ and $69$.  For all cases as the number of DMD modes included increases ($\gamma$ decreases), the residual generally decreases until a minimum value is reached before all DMD modes are included.  For all cases except the $S=5$ case, the residual becomes less than 5\% at the minimum or optimal number of DMD modes.  Interestingly, for the $S=69$ case, the minimum error is higher than minima of lower dimensionality reduction cases. 
(ii.) The L2 residual with respect to the difference between the sum over all mode pairs of scale-specific CKE and the TKE in the domain is given by
\begin{equation}
   \epsilon_{k} = \frac{\lVert k - \sum_{k=0,m=0}^{R,R} \frac{1}{2}\overline{\mu^k\mu^m}\alpha^k \phi^k_i \alpha^m\phi^m_i \rVert_2} {\lVert k \rVert_2},
   \label{eqn:k_residual}
\end{equation}
where all pairs of DMD modes are summed to reconstruct the CKE based on the DMD modes.  This metric is use to quantify the amount of kinetic energy  present in coherent scales verse random scales.  The closer the residual is to zero, the more kinetic energy is capture in the DMD modes and coherent scales.   In particular, we employ this residual metric to quantify the amount of TKE that is present in the coherent scales designated by DMD modes.
Similar to the reconstruction residual, all cases except the $S=5$ case, have a minimum residual error less than 5\%.   However, for all $S$ values, almost all $\gamma$ values select sets of modes that contain a majority of the total TKE.  The only values of $\gamma$ that do not have relatively high values is where no modes are select through compressive sensing.  Examples of these cases are shown in Fig. \ref{fig:selection}(b) for the $S=31$ case. In this example, as $\gamma$ decreases from the maximum, the smallest set of modes selected is $R=10$.  Similar behaviour is present for all values of $S$.

\begin{figure}
   \begin{center}
       \includegraphics[width=\textwidth]{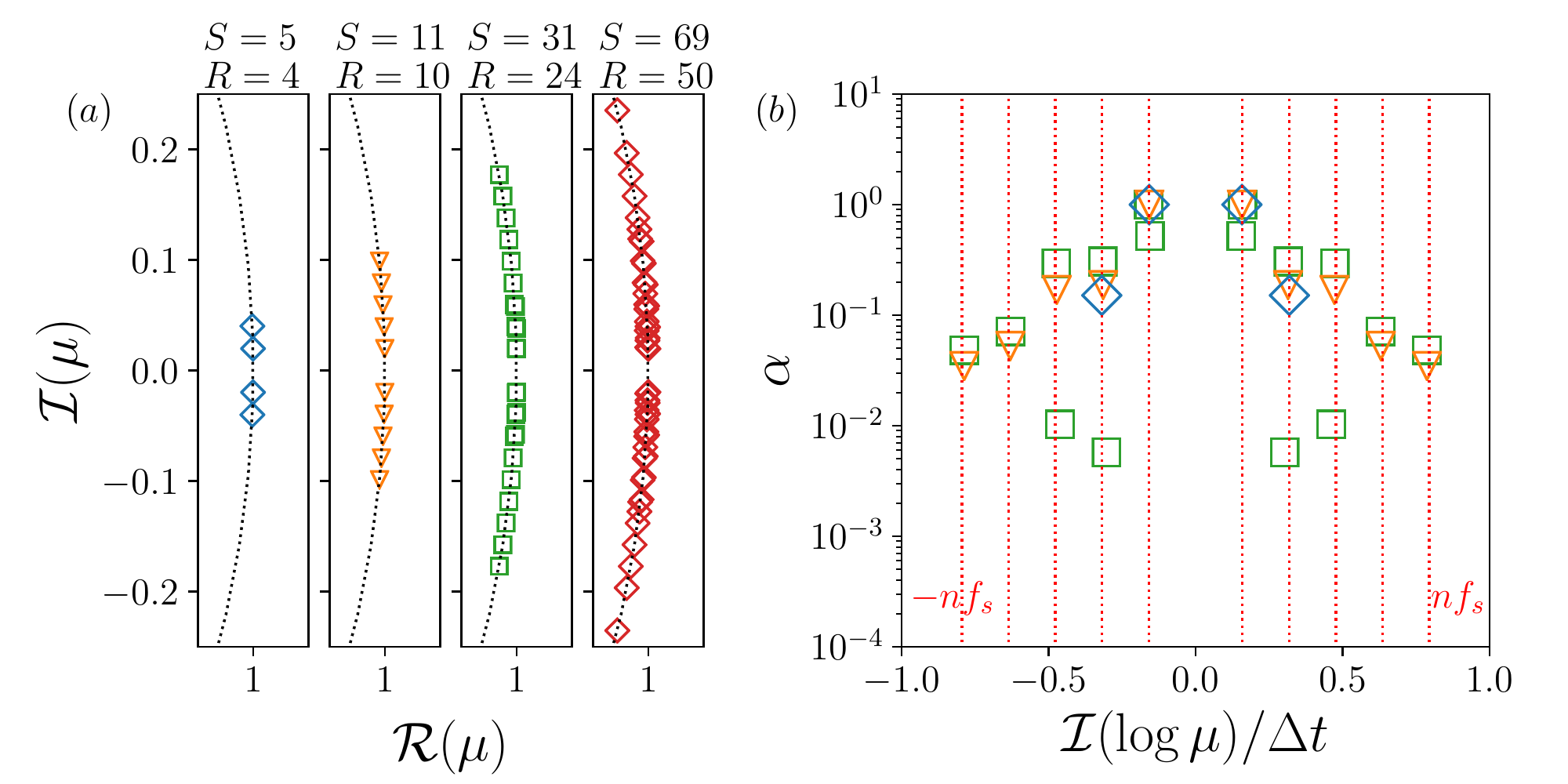}
       \caption{\label{fig:freq} (a) Ritz values $\mu_i$ for $S=5, 11, 31$, and $69$ and (b) amplitudes $|\alpha_i|/\max(\alpha)$ verse Strouhal number. The vertical dotted red lines identify integer multiples of the shedding frequency.}
    \end{center}
\end{figure}

Figure \ref{fig:freq}(a) shows the Ritz values $\mu_i$ on the real-imaginary plane for the $S=5, 11, 31$, and $69$.  The number of modes, $R$, select is based on the minimum $\epsilon_u$ residual for each $S$.  As the number of POD modes employed increases, additional DMD modes with Ritz values with a larger imaginary component, and their complex conjugates, are selected.  The imaginary component of the Ritz value is related to the non-dimensional frequency of each DMD mode: $\mathrm{St} = \mathcal{I}(\log \mu_i)/\Delta t (D/U_\infty)$. For the $S=31$ and $69$ case, some of the modes are represented by frequencies with approximately the same value. This can be seen clearly in the spectra of the DMD modes in Fig. \ref{fig:freq}(b), where the frequency of each selected DMD mode is shown and is related to harmonic multiples of the shedding frequency $f_s$. The relative strength of each mode in the DMD mode set is also shown in Fig. \ref{fig:freq}(b). For each $S$ value selected, there is at least one DMD mode that characterizes the shedding frequency, and one of these modes has the highest amplitude.  This demonstrates that this is most dominant mode in the set. For the $S=31$ case, there are two modes that have frequencies near the shedding frequency as well as the second and third integer multiples of the shedding frequency.  Furthermore, the DMD spectra confirm that the DMD algorithm is able to identify modes that are based on frequencies and corroborates the spectral energy signatures in Fig. \ref{fig:spectra}.  The spectra show that as more DMD modes are selected, the number of shedding frequency integer multiples represented increases. 

\begin{figure}
   \begin{center}
       \includegraphics[width=\textwidth]{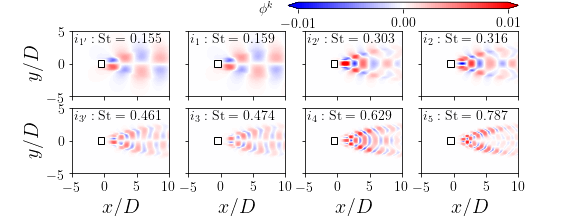}
       \caption{\label{fig:modes} DMD modes with the eight highest amplitudes from the $S=31, R=24$ and characterized by a positive frequency. }
    \end{center}
\end{figure}

The spatial DMD modes do not change significantly with large $S$ values because a substantial amount of energy is already accumulated with even the lowest $S=5$ value.  For example, the DMD mode for the shedding frequency, $i_1$, is the same for all $S$ analysed. However, the $S=5$ case only consists of the modes related the $i_1$ and $i_2$ frequencies (and their complex conjugates). The energy associated with any higher frequency modes that are not distinguished within a set can only be accounted for in the RKE.  
The DMD modes for the streamwise velocity associated with positive increasing Strouhal numbers are shown in Fig. \ref{fig:modes} for the $S=31$ case.   As expected, the mode associated with the shedding frequency consists of alternating pairs in the wake behind the cylinder.  For the $S=31$ case, a second mode associated with a frequency near the shedding frequency, $i_1^{\prime}$ has a similar structure.   Modes associated with higher multiples of the shedding frequency consist of additional alternating patterns.  Both the $i_2$ and $i_3$ DMD modes are associated with are DMD modes that have slightly a lower frequency.  These modes are attributed with a lower amplitude as shown in Fig. \ref{fig:freq}(b).  

\begin{figure}
   \begin{center}
       \includegraphics[width=\textwidth]{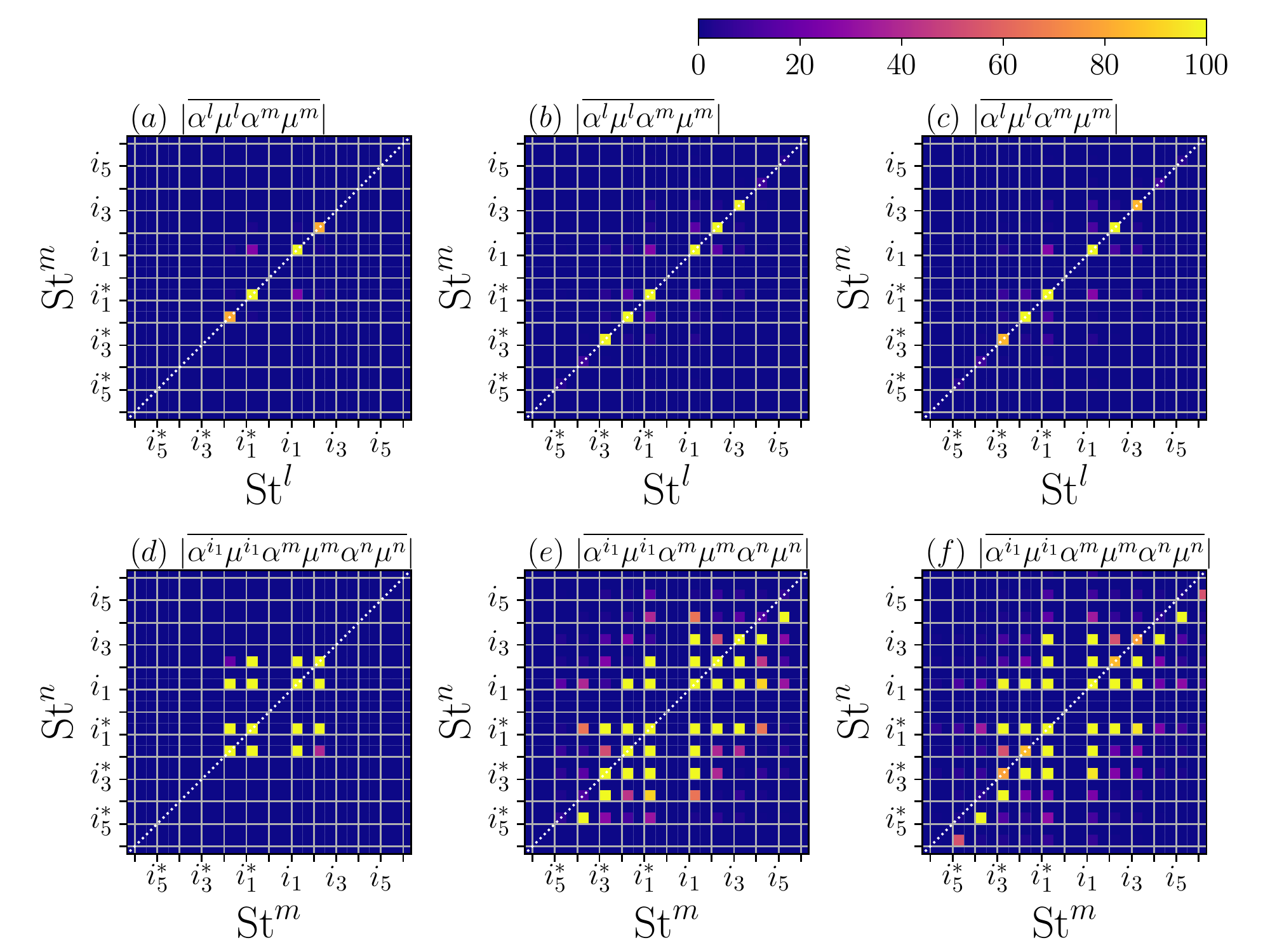}
       \caption{\label{fig:corr} The (top) dyadic and (bottom) triadic (with $l=i_1$) correlation between DMD modes for the (a),(d) $S=5$, (b),(e) $S=11$, and (c),(f) $S=31$.}
    \end{center}
\end{figure}

The relationship between scales can be quantified by the correlation of dynamic mode temporal coefficients and amplitudes.  The correlations appear as coefficients in all the terms in Eqn. (\ref{eqn:short_cke_ss}) and determine the overall magnitude of each term. 
The dyadic correlation of two dynamic modes appear in terms pertaining to the CKE transport via viscosity and pressure as well as the scale-specific CKE in Eqn. (\ref{eqn:ss_cke}), where it identifies the dominant mode pairs in CKE.  
The absolute value of the dyadic correlations, $\overline{\alpha^l\mu^l\alpha^m\mu^m}$, is shown in Figs. \ref{fig:corr}(a), (b), and (c) for the $S=5, 11$ and $31$ cases, respectively.  While it reveals that contribution of the scale to $\overline{\tilde{k}}^{l,m}, \mathcal{T}_p^{l,m}$, etc, it is dominated by only a small fraction of the selected modes.   The high correlation contributions are aligned along the $\mathrm{St}^l = \mathrm{St}^m$ line, which indicate that the highest correlations are between pairs with $l=m$.  In fact, the highest correlation for each case is $i_1 = i_1$ case and its complex conjugate ($i_1^* = i_1^*$).  The next highest correlations are associated with the $i_2$ frequencies. Each case also displays at least one off-diagonal case, which is the correlation between the conjugate-pairs.   This demonstrates for this flow that the dominant scale-specific CKE scales are the ones that consist only one scale, i.e., $\overline{\tilde{k}}^{i_1,i_1}$, and suggest that energy exchange between scales is not quantified by dyadic correlations. 

However, the triadic interactions show that much more diverse and less sparse sets of modes are correlated in Figs. \ref{fig:corr}(d), (e), and (f) showing the triadic correlations for the $S=5, 11$ and $31$ cases, respectively.
The triadic correlations appear in the turbulent transport $\mathcal{T}_t^{l,m,n}$ and inter-scale transfer $\mathcal{P}_t^{l,m,n}$ as $\overline{\alpha^l\mu^l\alpha^m\mu^m\alpha^n\mu^n}$. Figures \ref{fig:corr}(d), (e), and (f) shows the correlation of triadic interactions of $l$th mode, where $l=i_1$.  Each $S$ case demonstrates that triadic interactions with high correlations are distributed throughout the frequency space. Not only are there high correlations along the diagonal $\mathrm{St}^m=\mathrm{St}^n$, but off-diagonal correlations reveal that turbulent transport and inter-scale transfer occur over disparate scales.  The $S=5$ case is symmetric around the diagonal, while the sets that have  more modes and contain more of the total TKE are asymmetric. Higher correlations occur off-diagonal where $\mathrm{St}^m$ is large.  The $m$th contribution to the inter-scale transfer, $\mathcal{P}_t^{l,m,n}$ is the mode where CKE is remove and placed in the CKE related to $l$ and $n$.  A high correlation demonstrates that CKE transfer from/to the $m$th is high.  This will be more thoroughly analysed below.   Figures \ref{fig:corr}(d), (e), and (f) indicate that interactions between scales is primarily accomplished through triads.  
On the other hand, the dominant scale-specific kinetic energy and scale-specific dyadic terms in Eqn. (\ref{eqn:short_cke_ss}) consist of one scale. This confirms that CKE contributions are based on a single scale, but the exchange of CKE between the scale is requires multiple scales. 

\begin{figure}
   \begin{center}
       \includegraphics[width=\textwidth]{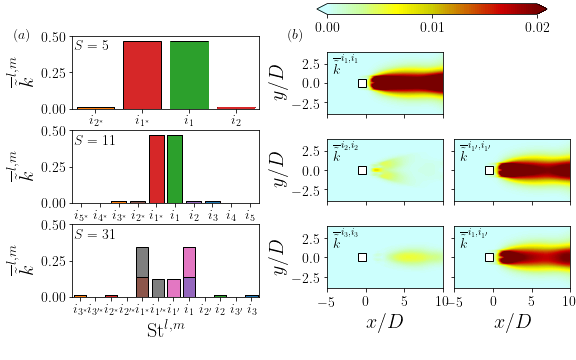}
       \caption{\label{fig:tke_frac} (a) The fraction of scale-specific CKE to the total CKE for the $S=5, 11$ and $31$ case and (b) spatial distribution of scale-specific CKE. }
    \end{center}
\end{figure}

We now focus on the attributes of the scale-specific CKE and the relative contributions from each DMD pair.  Figure \ref{fig:tke_frac}(a) shows the fractional contribution from each mode pair to the total CKE for the $S=5, 11$ and $31$ cases.  Bars that have a range starting at zero represent the contributions from the self-interactions of a mode e.g., $(i_1,i_1)$. Bars that stacked on other bars represent the contributions from two different modes pair e.g., $(i_1,i_1^\prime)$. The stacked bar colouring matches the colour of another bar, where the range starts at zero, to show which two, different modes are represented.  This only occurs in the $S=31$ case, where more modes are identified with different features.   The main contribution of CKE comes from the modes with frequencies closest to the shedding frequency, $i_1$ and $i_{1^\prime}$ and their complex conjugates: $i_1^*$ and $i_{1^{*\prime}}$. Together the interaction among these four modes totals over 94\% of the CKE for each $S$ value.  Also, note that the scale-specific CKE contributions for a mode and its complex conjugate are exactly the same because $\overline{\tilde{k}}^{i_1,i_1} = \overline{\tilde{k}}^{i_1^*, i_1^*}$.  The remainder of the CKE contribution come from the $i_2$ and $i_2^*$ modes for the $S=5$ case.  In the other two cases, the contributions come from both the second and third integer multiples of the shedding frequency.   However, 
these only make up approximately 5.9\% of the total CKE.  The remaining 0.1\% of the CKE is associated with the rest of the mode self-interactions at higher multiples for the $S=11$ and $S=31$ cases.  Overall, the modes related to the shedding frequency have the dominant impact on the flow and contribute the most to the CKE.  This can be observed in Fig. \ref{fig:tke_frac}(b) where the scale-specific CKE for the $(i_1,i_1)$ pair looks strikingly similar to the TKE in Fig. \ref{fig:velocity}(c).  For the $S=31$ case, the scale-specific CKE associated with $i_1^\prime$ are shown in Fig. \ref{fig:tke_frac}(b) and have a similar spatial distribution as the CKE as well.  Also, each conjugate-pair scale-specific CKE have the same spatial distribution.  The scale-specific CKE related the  second  $(i_2, i_2)$ and third $(i_3,i_3)$ multiples have a distribution near and far downstream, respectively, of the cylinder. This suggests that energy is distributed in higher frequency modes at different downstream locations. 

\begin{figure}
   \begin{center}
       \includegraphics[width=\textwidth]{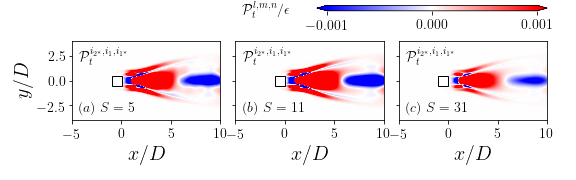}
       \caption{\label{fig:ptijk_mode2} Spatial distribution of the inter-scale transfer from mode $i_1$, $\mathcal{P}_t^{i_2,i_1,i_1}$, normalized by turbulent dissipation $\epsilon$ for the (a) $S=5$, (b) $S=11$, and (c) $31$ cases.}
    \end{center}
\end{figure}

The inter-scale transfer $\mathcal{P}_t$ is constructed from the sum of all triadic scale-specific interactions. The largest scale-specific inter-scale transfer for all three $S$ values is associated with the triadic interactions of the first and second integer multiples of the shedding frequency, $i_1$ and $i_2$, respectively. Figure \ref{fig:ptijk_mode2} shows the spatial distribution of the inter-scale transfer.  By convention, positive values indicate transfer from the $i_1$ to the correlation of $i_1$ and $i_2$, i.e., $\mathcal{P}_t^{i_2,i_1,i_1} = \overline{\tilde{u}_i^{i_1}\tilde{u}_j^{i_2} \partial_j \tilde{u}_i^{i_1}}$.  The transfer of CKE from the most dominant coherent structure to other modes occurs in the wake of the cylinder and along the shear layer of the expanding wake.   There is reverse transfer in the far wake ($x/D>5$) and immediately behind the cylinder. Each case is able to identify the same dominant triadic relationship.    

\begin{figure}
   \begin{center}
       \includegraphics[width=\textwidth]{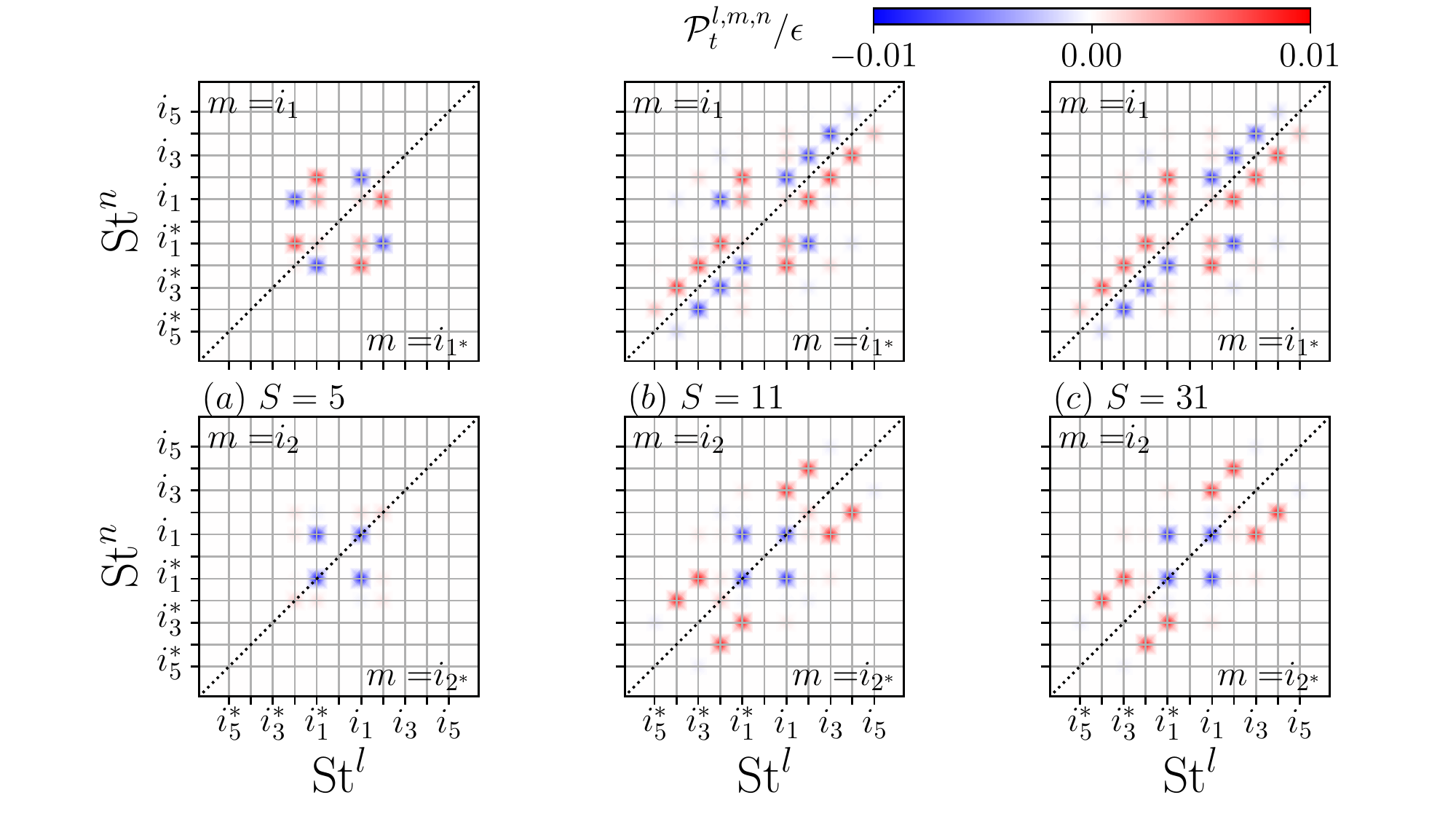}
       \caption{\label{fig:ptijk_int2} The magnitude of the scale-specific inter-scale transfer $\mathcal{P}_t^{l,m,n}$ at $x/D=1$ and $y/D=0$ for the (top) $m=i_1,i_1^*$ and (bottom) $m=i_2,i_2^*$ with (a) $S=5$, (b) $S=11$, and (c) $S=31$.}
    \end{center}
\end{figure}

The magnitude of the scale-specific inter-scale transfer associated with the transfer of kinetic energy from $i_1$ and $i_2$ at $x/D=1$ and $y/D=0$ is shown in Fig. \ref{fig:ptijk_int2} for each $S$ value. The inter-scale transfer from the $m$th mode is designated by detailing the magnitude of the transfer to the $l$th and $n$th modes.    Starting with $S=5$ in Fig. \ref{fig:ptijk_int2}(a), there are only two conjugate pairs of modes that can exchange CKE, however, there are a total of 64 triadic interactions that are possible. Each scale-specific inter-scale transfer is symmetric across the $\mathrm{St}^l$ = $\mathrm{St}^n$ line where $m$ is above the line and its complex conjugate $m^*$ is below the line.   For the $m=i_1,i_1^*$ inter-scale transfer, there is both inter-scale transfer from and to shedding frequency.  All consist of the $i_1$-$i_2$ pair, which indicate that energy is being exchanged due to a strong correlation in CKE between the two frequency modes.  On the other hand, the $i_2,i_2^*$ inter-scale transfer reveals that CKE is only transferred from $i_1$ to $i_2$ and the $i_2$ does not have a strong transfer to the shedding frequency.  

The number of triadic interactions leading to inter-scale transfer increases for the $S=11$ and $S=31$ cases shown in Fig. \ref{fig:ptijk_int2}(b) and (c), respectively.  The inter-scale transfer at the $x/D=1$ along the centerline is similar for both cases.  In the $m=i_1,i^*_1$ inter-scale transfer, CKE is symmetrically transferred to other modes that are associated with higher integral multiples of the shedding frequency. However, the energy is transferred back to this mode.  The $m=i_2,i_2^*$ transfer shows a cascade-like energy exchange where CKE is transferred from the $i_1$ shedding frequency and then transferred to higher frequencies.  This indicates that CKE transfers follows different paths, one from the shedding frequency directly and another from the lower frequency modes.  In the latter, the cascade to higher frequencies continues when $m>i_2$.  Both the $S$ values show similar behaviour for the inter-scale transfer at low frequencies where DMD was able to identify them. However, the $S=31$ case captures more of the high frequency behavior and can identity the inter-scale relationship of those modes, too. 

\begin{figure}
   \begin{center}
       \includegraphics[width=\textwidth]{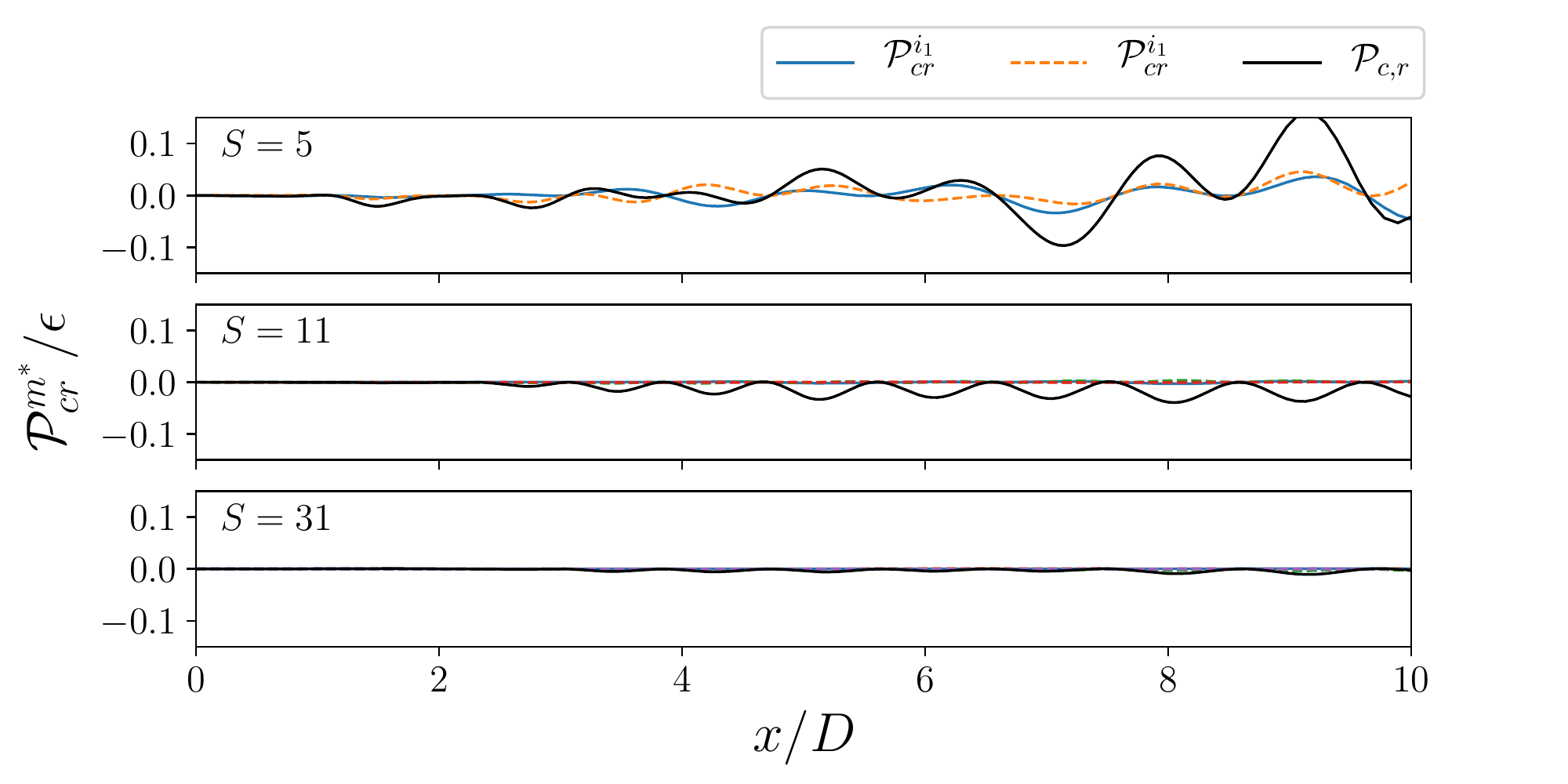}
       \caption{\label{fig:prc_centerine} The scale-specific random production from coherent scales $\mathcal{P}_{cr}^{m*}$ and total random production from coherent scales $\mathcal{P}_{cr}$ for the (a) $S=5$, (b) $S=11$, and (c) $S=31$. }
    \end{center}
\end{figure}

Another inter-scale transfer is the random production from coherent scales term $\mathcal{P}_{cr}$, which capture the transfer of CKE to RKE.  The developed methodology is able to identify how each coherent scale transfer energy to balance the evolution of RKE.  Figure \ref{fig:prc_centerine} shows the two largest mode contributors for $S=5, 11$ and $31$ as well as the total $\mathcal{P}_{cr}$.  In all cases the largest contributing scales are the scales related to the $i_1$ and $i_2$ scales.  The sum of the contributions from these two modes contribute of 95\% of the total random production from coherent scales for all cases.  However, as the number of identified modes increases ($S$ increases), the total amount of random production from coherent scales decreases significantly.  This is because more scales have been identified and less RKE is present.  Overall, the production term indicates that transfer of energy is linked to the shedding frequency. 

Based on the two error residuals and subsequent analysis, there are some differences in how to select the appropriate number of modes.  As the total TKE included in the modes increases, the relationships between the triadic interactions and inter-scale production begin to converge.  While there may exist some heuristic to identify an optimal number of modes to select, we will use subjective judgement based on the number of modes and the amount of TKE that can be identified as CKE.   For this case, we observe that the $S=31$ case (the total cumulative energy is 99.99\%) has DMD solutions that are less than 2\% for both residual metrics. The minimum reconstruction residual with $R=24$ is less than 1\% and has a similar TKE residual at higher values of $R$.  The main difference between this case, and the $S=11$ case is the presence of many higher frequencies that can be used to identify the energy cascade-like features in the inter-scale energy transfer.   In what follows, we will use the particular set of DMD modes: $S=31$ , $R=24$.

\begin{figure}
   \begin{center}
       \includegraphics[width=\textwidth]{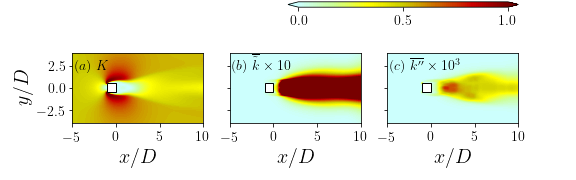}
       \caption{\label{fig:triple} Contours of the  (a) MKE, (b) CKE, and (c) RKE for the $S=31$ case. }
    \end{center}
\end{figure}

Figure \ref{fig:triple} details the spatial distribution of the components of the kinetic energy due to the triple decomposition.  The MKE in Fig. \ref{fig:triple}(a) is relatively high in all regions of the domain except near the stagnation point where the velocity is decreases to zero and in the wake of the cylinder from $0 \le x/D \le 4$.  The MKE in the wake gradually recovers as the wake expands and entrains MKE into the wake.  The MKE is highest at the leading edge of the square cylinder where the mean velocity increases due to the blockage of the cylinder.  The total CKE, shown in Fig. \ref{fig:triple}(b) shows energy associated with coherent motion is spatially distributed in the wake of the cylinder and is closely associated with the TKE shown in Fig. \ref{fig:velocity}(c). This is mainly due to our selection of DMD modes where almost all the TKE is accounted. On the other hand, Fig. \ref{fig:triple}(c) shows the RKE, which is the residual kinetic energy that is not accounted for through mean or coherent motion. While the RKE is only located in the wake, similar to the CKE, it is several orders of magnitude smaller that the MKE and CKE.  This is the energy that is associated with all modes that are not selected by sparse sampling and dimensionality reduction.  

%
\subsection{Inter-scale transfer and interactions of coherent scales}
The mode decomposition used to create the scale-specific inter-scale transfer $\mathcal{P}_t$ provides a clear link to the non-linear and non-local interaction of scales.  The inter-scale transfer is the only term in the scale-specific CKE budget, Eqn. (\ref{eqn:ss_cke}), that is a function of three modes, is based on their associated frequencies, and links the scale-specific CKE equations together.  The term appears as a turbulent production-like term where the gradient of one scale transfers CKE to another scale. Furthermore, similarity to the Fourier transform of the energy equation can be established between the presences of triadic interactions. In our context, the mode decomposition of the inter-scale transfer also links a triadic relationship with identified components and spatial fluxes in the flow.  In the context of the Richardson-Kolmogorov equilibrium, the inter-scale transfer identifies how specific modes of energy interact and transferred.

\begin{figure}
   \begin{center}
       \includegraphics[width=.8\textwidth]{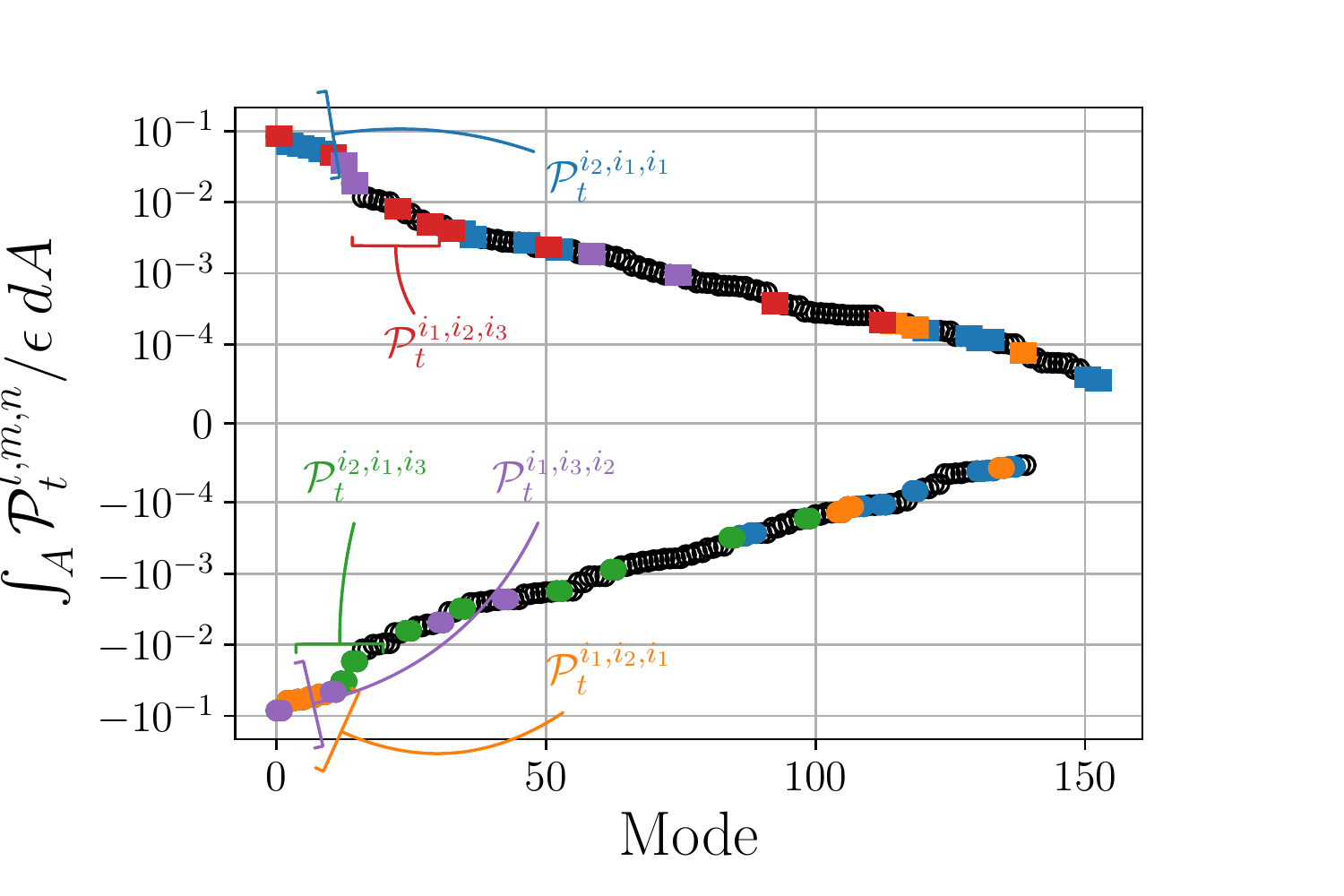}
       \caption{\label{fig:ptijk_spectra} The area integral of the scale-specific inter-scale transfer $\int_A \mathcal{P}_t^{l,m,n} dA$ normalized by the integral turbulence dissipation $\epsilon$.  Select classes are identified by colour.}
   \end{center}
\end{figure}

We first concentrate on identifying the dominant triadic contributions to the inter-scale transfer by assessing which triads have the highest fraction of the total inter-scale transfer.  The contribution is assessed by integrating the scale-specific inter-scale transfer over the entire spatial domain: $\int_A \mathcal{P}_t^{l,m,n} dA$.  The sum of all integrated terms is equal to the spatial integral of the total inter-scale transfer.  Figure \ref{fig:ptijk_spectra} shows the fractional proportion triadic interactions in the inter-scale flux normalized by the turbulent dissipation $\epsilon$.  The contributions create a dual spectrum where some terms are negative, indicating transfer to the $m$th, while the majority are positive, indicating transfer from the $m$th scale.  The spectrum also reveals that there is a large separation between the most dominant and smallest contributions on each side of zero.  About 10\% of the total number triads on each side of the spectrum contribute over 80\% of the total inter-scale transfer. Of these largest relative contributions, several classes of triadic relationship are identified on Fig. \ref{fig:ptijk_spectra}.  
Each class consists of the interaction of three modes, where the $m$th mode can either be $m$ or its complex conjugate $m^*$. The $l$th and $n$th modes in each class is either another frequency or its conjugate. The spatial distributions of the classes are degenerates such that they have the same, but scaled distribution.   The possible permutations mean that each class can contain at least 64 triadic interactions.  The largest contributions on the positive side are associated with $m=i_1,i_1^*$ where $l$ and $n$ are $i_1$ and $i_2$. This indicates that overall CKE is transferred from the shedding frequency to the CKE associated with the relationship of the shedding frequency and its second integer multiple. CKE is transferred from a low, dominant frequency mode to a mode associated with higher frequency.   Another dominant contribution on the positive side of the spectrum is the class of $(l,m,n) = (i_1, i_2, i_3), (i_3, i_2, i_1^*)$, etc.,  where CKE from second dominant mode, $i_2$ is transferred to CKE associated with the first and third integer multiple of the shedding frequency.  Each triadic class on the positive side of the spectrum exhibits similar behaviour: the $m$th mode has a lower frequency and at least one of the $l$ or $n$th modes is associated with a higher frequency.
These two sets of mode interactions indicate a transfer of energy between the first, second, and third integer multiple of the shedding frequency.  They imply that interactions between the vortex shedding transfers energy from modes associated with the shedding frequency to modes related to higher harmonics.

The negative side of the spectrum also reveals classes of triad that promote the cascade of CKE from low to high frequencies. These includes one of the dominant classes of $(l,m,n) = (i_1, i_2, i_1$, which identifies another pathway for CKE to transfer from the coherent motions of the shedding frequency to its second integer multiple.  The cascade is continued on this side of the spectrum with another class of triads: $(l,m,n) = (i_1,i_3,i_2)$, where energy is transferred to the third integer multiple.  
The spectrum in Fig. \ref{fig:ptijk_spectra} also reveal there is inverse transfer of kinetic energy from high frequencies to lower frequencies.  CKE is transferred from the energy in the correlation of the $i_2$ and $i_3$ modes to the $i_1$ mode. This class of triads represents a significant portion of the total inter-scale transfer.  
Overall, the interactions all occurs with  the shedding frequency, showing the non-linearity and non-locality that is present in wake.   

\begin{figure}
   \begin{center}
       \includegraphics[width=\textwidth]{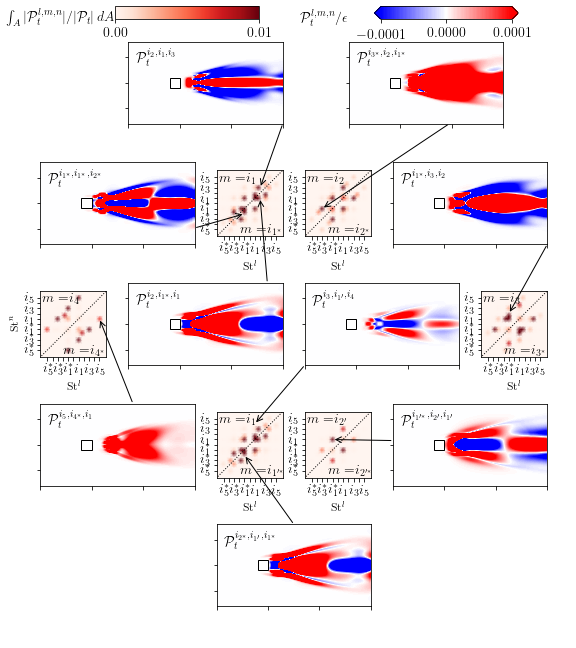}
       \caption{\label{fig:ptijk_modes} Spatial distribution of the scale-specific inter-scale transfer $\mathcal{P}_t^{l,m,n}$ of select triads.  Contours of the fractional proportion of the total inter-scale transfer $\int_A |\mathcal{P}_t^{l,m,n}|/|\mathcal{P}_t| \: dA $ is shown in frequency space $\mathrm{St}^l \times \mathrm{St}^n$ for $l=i_1, i_{1^\prime}, i_2, i_{2^\prime}, i_3$ and $i_4$.}
   \end{center}
\end{figure}

The spatial distribution of the $\mathcal{P}_t^{l,m,n}$ identifies how inter-scale transfer behaves in the wake.  Figure \ref{fig:ptijk_modes} shows several examples of the modes for specific triads that are representative of their class.  Additionally, the fractional contribution of the total inter-scale transfer is shown as series of contours in frequency space as $\mathrm{St}^l \times \mathrm{St}^n$ for a specific $m = i_1, i_{1^\prime}, i_2, i_{2^\prime}, i_3$ and $i_4$ and their complex conjugates.  The frequencies and their complex conjugates are symmetric around the $\mathrm{St}^l = \mathrm{St}^n$ line, which provides further indication that the triadic set of modes is comprised of permutations of frequencies in a  triadic set.  Their scale is related to the order of the $l$ and $n$ as well as whether the mode is a complex conjugate.  The spatial distribution is also based on the order as well.  

The fractional distribution of the terms where $m=i_1, i^*_1$ confirms that there are multiple dominant triads involving the shedding frequency. As such there are multiple paths for CKE to transfer from energy in the vortex shedding to other scales. The largest positive scale-specific inter-scale transfer as shown in Fig. \ref{fig:ptijk_spectra} is the class of triads $(l,m,n) = (i_1,i_1,i_2)$.  The spatial distributions of $P_t^{i_1^*,i_1^*,i_2^*}$ and $P_t^{i_2,i_1,i_1}$ are shown in Fig. \ref{fig:ptijk_modes}.  In both there are regions of both positive and negative inter-scale transfer.   The regions of positive inter-scale transfer occur along the shear layer where vortex shedding and the wake grow. In these regions CKE is transferred from the $i_1$ mode and transferred to $i_2$.   The wakes at $x/D>5$ differ substantially between the two cases.   Furthermore, along the centerline, the transfer transitions from positive to negative.  Another dominant triad involved in inter-scale transfer is the $(l,m,n) = (i_2,i_1,i_3)$, which shows a different spatial distribution related to the transfer CKE from $i_1$.  The CKE transfer transitions from positive to negative in the shear layers and is positive along the centerline.  
These triads incorporate the exchange of energy from the mode associated with the shedding frequency.  Other similar classes of triads incorporate the transfer from the $m=i_{1^\prime}i_{1^\prime}^*$ frequency.  Due to the proximity to the shedding frequency, the spatial distributions of inter-scale transfer associated with these triads are similar to inter-scale transfer from $m=i_1,i^*_1$.  In fact, the distribution of fractional contributions of the $m=i_{1^\prime}i_{1^\prime}^*$ frequencies is very similar to the  $m=i_1,i^*_1$ indicating that CKE in this frequency is also distributed to many of the other scales.   One triad shown $(l,m,n) = (i_3,i_{1^\prime},i_4)$ exhibits energy transfer to very high frequency modes and captures the non-local effects of the inter-scale transfer.  Overall, the effects of the triads with $m=i_{1^\prime}$ frequencies contribute less to the total inter-scale transfer than the $m=i_1$ modes.  

Other triads that transfer CKE from moderate to high frequency modes are also shown in Fig. \ref{fig:ptijk_modes}.  The triad $(l,m,n) = (i_3,i_2,i_1^*)$  has a spatial distribution that is mainly positive throughout the wake.  Similarly, another triad $(l,m,n) = (i_5,i_4^*,i_1)$, transfers CKE to higher frequencies.  Different from the other spatial distributions, the inter-scale transfer with $(l,m,n) = (i_1^*,i_3,i_2)$ demonstrates that high frequency scales can also transfer energy to lower frequencies.  The $m=i_{2^\prime}$ cases have lower contributions to only a few other scales.  Overall, most of the dominate inter-scale transfer terms involve a triad with $i_1,i_1^*$.  The CKE in this scale dominants the flow and provides the mechanism to transfer the energy to the other scales.
Another observation from the contours of the fraction of total inter-scale transport that is that as $m$ increases, the regions of high contribution move outward from the centre. There is a  higher probability for high transfer to occurs with higher frequencies. However, because the shedding frequency is so dominant in this case, the contributions regardless of the $m$ are always likely contain the shedding frequency.

\begin{figure}
   \begin{center}
       \includegraphics[width=\textwidth]{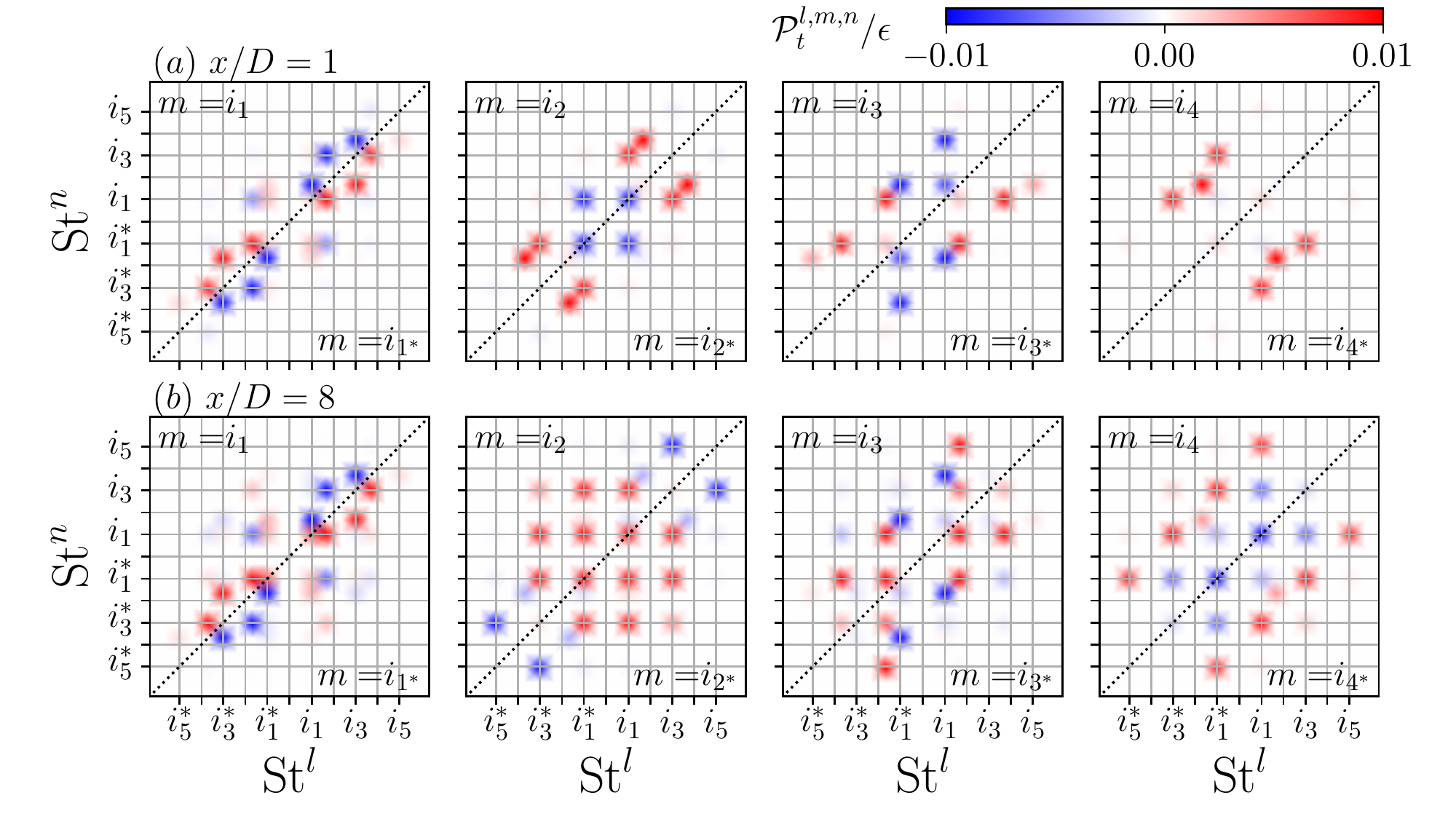}
       \caption{\label{fig:ptijk_interact} Spectra of the scale-specific inter-scale transfer $\mathcal{P}_t^{l,m,n}$ of $m=i_1, i_2, i_3$ and $i_4$ normalized by the total turbulence dissipation at (a) $x/D=1$ and (b) $x/D=8$ along the centerline.}
   \end{center}
\end{figure}

The spectra of the $m$th scale-specific inter-scale transfer at a single location in the wake is shown in Fig. \ref{fig:ptijk_interact} to demonstrate the relationships among triads and identify strong interactions of the scales in the wake.  The two-dimensional spectrum of the inter-scale transfer from the $m$th mode to modes associated with frequencies across the frequency space, $\mathrm{St}^l \times \mathrm{St}^n$, at $x/D=1$ along the centerline is shown in Fig. \ref{fig:ptijk_interact}(a) for $m=i_1, i_2, i_3$ and $i_4$. The contributions of each $m$th mode is anti-symmetric around the line $\mathrm{St}^l = \mathrm{St}^n$ line, where contributions from the $m$th mode and its complex conjugate $m^*$ are above and below the line, respectively.  The contours allow us to interpret the exchange of CKE between the triads.  For instance, when $m=i_1$ at $x/D=1$, CKE is transferred to the component of the scales with the correlation $(l,n) = (i_2^*, i^*_1)$, while energy is transferred from scales $(l,n) = (i_1, i_2)$.  The scale-specific inter-scale transfer from $m=i_1,i^*_1$ is both positive and negative at $x/D=1$ indicating that CKE is transferred to and from the shedding frequency mode at similar rates.  This cyclic inter-scale transfer trend occurs at higher frequencies that extend along the $\mathrm{St}^l = \mathrm{St}^n$.  This indicates that the $m=i_1, i^*_1$ modes transfer CKE to $l,n$ scales that are associated with the same or similar frequencies or are not complex conjugates.   

The inter-scale transfer modes along the centerline at $x/D=1$ changes with increasing frequency of the $m$th modes. For the  $m=i_2,i^*_2$ modes, Fig. \ref{fig:ptijk_interact}(a) shows that CKE is transferred to the second multiple of the shedding frequency by modes related to the shedding frequency.  This shows CKE in the near wake is transferred from the largest vortex shedding frequency to harmonics of the shedding near the genesis of the shedding.  Furthermore, the $m=i_2,i^*_2$ modes transfer CKE to higher frequencies.  The inter-scale transfer captures this distribution process to the high frequency modes.  This inter-scale processes enables the vortex shedding to distribute energy to multiple scales in the wake.   A similar process is captured by the $m=i_3,i_3^*$ modes, which receive energy from low frequencies and transfer it to higher frequencies.  However, we observe that at this higher frequency there is a feedback to the $i_1,i^*_1$ mode, which is present in the correlation as $l$th or $n$th mode in the spectra.   On the other hand, the $m=i_4,i^*_4$ modes are all positive indicating a positive transfer of energy to lower frequency modes.  Due to the low Reynolds number nature of the flow, these dominant modes demonstrate significant energy transfer to the energy range of dominant frequencies.   

The inter-scale transfer changes significantly in the far wake  at $x/D=8$ as shown in Fig. \ref{fig:ptijk_interact}(b).  Overall, there is a larger range of frequencies that are incorporated in the inter-scale transfer at this location.  The energy in the wake has been developed and CKE has be distributed to more scales, which play a role in the transfer in the far wake. The dominant triads responsible for inter-scale transfer are anti-symmetric around the $\mathrm{St}^l = \mathrm{St}^n$ line. The mode associated with the shedding frequency has similar behaviour to the $x/D=1$ location where there is both transfer from and to specific modes where the sign is dependent on the sign of the frequency.  This suggests that the vortex shedding, which is still active in the far wake, is source and sink of most of the CKE and enables the transfer to other scales in the flow.  
However, as $m$ increases, transfers between scales becomes different than the upstream location.  The transfer of energy from the $m=i_2,i_2^*$  occurs over wider range of the frequencies, where the transfer of energy has reversed.  The highest frequency modes are  associated with transfers to this mode, while the low energy is transferred to this mode.  The $m=i_3,i^*_3$ inter-scale transfer is similar to the $x/D=1$ location except over a larger range.  The highest $m=i_4,i_4^*$ mode is shown to receive CKE from the low frequencies and transfer CKE to higher frequencies.   Note that even at higher frequencies inter-scale transfer, the shedding frequency plays a role in the transfer.  

\begin{figure}
   \begin{center}
       \includegraphics[width=\textwidth]{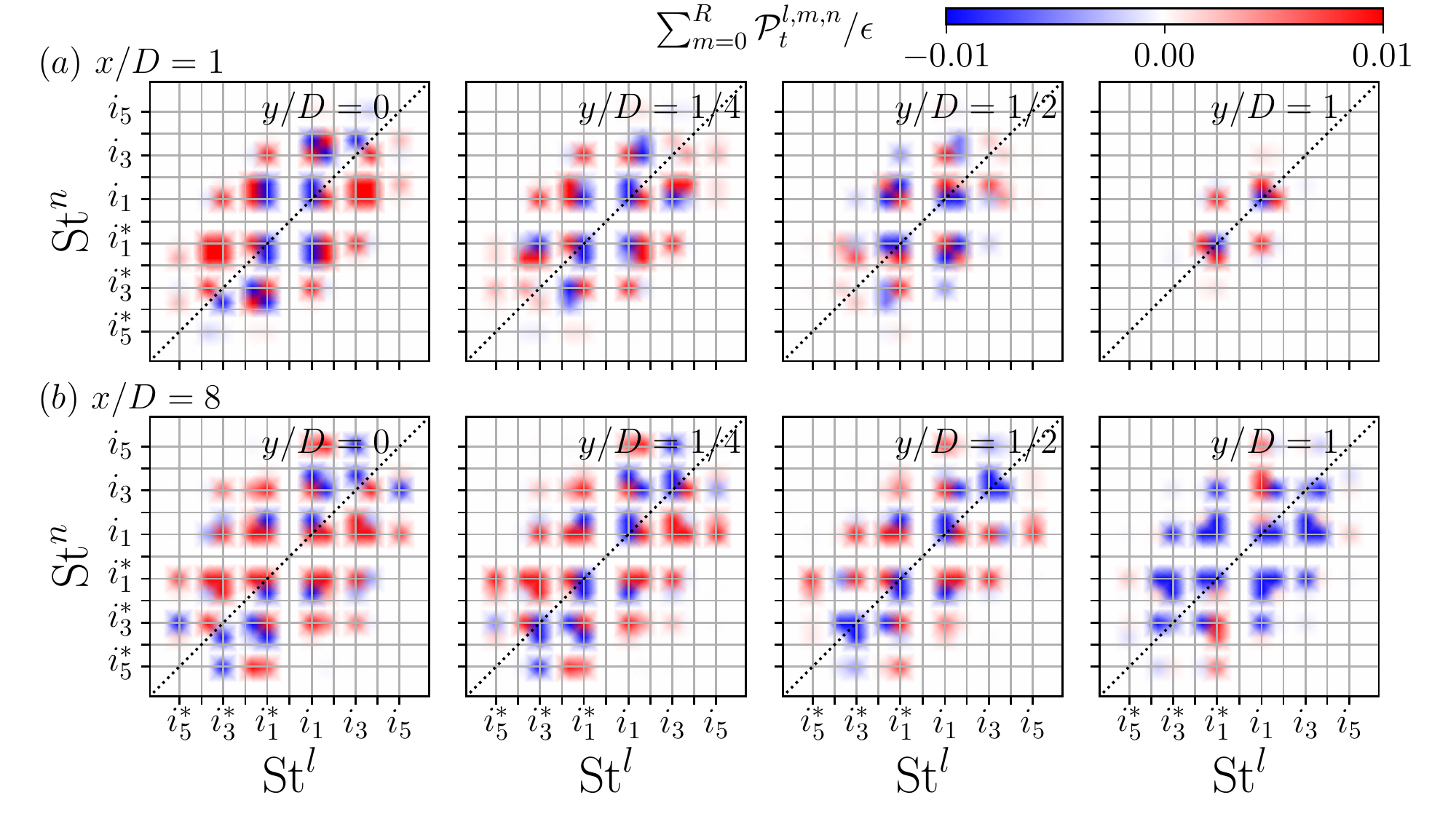}
       \caption{\label{fig:ptijk_interact_all} Spectra of the sum over $m$ scale-specific inter-scale transfer $\sum_{m=0}^R \mathcal{P}_t^{l,m,n}$ normalized by the total turbulence dissipation at $y/D=0, 0.25, 0.5$, and $1$ in the near wake (a) $x/D=1$ and far wake (b) $x/D=8$.}
   \end{center}
\end{figure}

The spectra of CKE inter-scale transfer over all $m$ modes, $\sum_{m=0}^R \mathcal{P}_t^{l,m,n}$, identifies the total energy transfer to/from all scales in the wake.  Figure \ref{fig:ptijk_interact_all}(a) shows the inter-scale transfer contributions for all $m$ modes at several spanwise locations, $y/D=0, 0.25, 0.5$, and $1$ in the near wake at $x/D=1$.  Similar to findings in Fig. \ref{fig:ptijk_interact}, the spectra are anti-symmetric around the line $\mathrm{St}^l = \mathrm{St}^n$, where contributions from  all $m$ modes and all complex conjugate $m^*$ modes are above and below the diagonal line, respectively. The distribution of the energy transfer to/from different components radically changes with increasing spanwise distance from the centerline.  At the $y/D=0$, modes with frequencies  $l,n=i_1,i^*_1$ are generally negative.  This is clear indication that the transfer of energy is in the direction from the scale of the vortex shedding that forms near this location.  Furthermore, the transfer from the shedding frequency scale is corroborated by previous spectra in Fig. \ref{fig:ptijk_interact}, where CKE is shown to transfer from the largest scale.   On the other hand, higher frequency modes in Fig. \ref{fig:ptijk_interact_all}(a) at $y/D=0$ are generally positive indicating modes associated with this frequency receive CKE from other modes.     
At small radial distance from the centerline at $y/D=0.25$, the interactions among the triads act similar to the centerline.  The main difference is that the magnitude of the inter-scale transfer is smaller.  The exchanges or transfers of energy from the vortex shedding leads to the creation of modes associated with the higher integer multiples of the shedding frequency.
At the larger distances from the centerline $y/D=0.5$ and $y/D=1$, the magnitude of the transfer decreases further due to its position outside of the wake.  Less CKE and its associated transfer and transport are created outside the wake of the cylinder.  Furthermore, the exchange of energy at these locations is dominated transfers of energy around modes associated with the shedding frequency. The transfer to higher frequencies is inhibited by the lower gradients of the modes at these locations.

Figure \ref{fig:ptijk_interact_all}(b) shows the spectra of the inter-scale transfer from all $m$ modes in the far wake at $x/D=8$ at several $y/D = 0, 0.25, 0.5$, and $1$ locations.  Three locations, $y/D = 0, 0.25$, and $0.5$ are located inside the wake of the cylinder and have relatively similar spectra.  Overall in the far wake, the transfer of energy occurs over a larger range of frequency modes compared to the near wake $x/D=1$ locations.  As the wake evolves downstream towards the far wake, more CKE can be transferred to higher frequencies.  Furthermore, the wake becomes more homogeneous as it transition from the near wake to the far wake.  This is due to the fact that is takes a finite time to transfer energy from the scales of the shedding vortex to higher scales.  At the $x/D=1$ locations, the energy is present in vortex shedding is just beginning to transfer to other modes.  As a consequence of the large range in the spectra at $x/D=8$, we see that there is significant energy transfer from high frequencies as well.  Outside the wake, at $y/D=1$, CKE transfer is mainly negative for all frequencies.  The effects here may be due to balance of CKE and lead to contributions to other terms in the CKE equation.   
\subsection{Contributions from different terms in CKE equation}
\begin{figure}   
   \begin{center}
       \includegraphics[width=\textwidth]{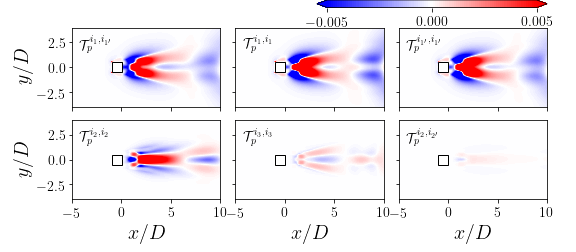}
       \caption{\label{fig:tp_top} Contours of six largest scale-specific contributions of the CKE transport via pressure.}
    \end{center}
\end{figure}

Next, we assess the other terms in the scale-specific CKE equation based on their modal contributions of pairs of modes.  The transport of total CKE is subject to three different phenomena: pressure, viscosity, and turbulence.  We start with the dominant CKE transport via pressure, $\mathcal{T}_p$.  The scale-specific CKE pressure transport is given in Eqn. (\ref{eqn:tp_ss}) and based on dyadic interactions of modes. The spatial distribution of the six largest contributions, $\mathcal{T}_p^{l,m}$, to the total $\mathcal{T}_p$ are shown in Fig. \ref{fig:tp_top}.  Similar to the scale-specific CKE, the largest contributions are from modes related to the shedding frequencies and their complex conjugates: $i_1$ and $i_{1^\prime}$.  The largest contribution is from the interaction of velocity and pressure fluctuations of the pair ($i_1, i_{1^\prime}$) and their complex conjugates.  Note that $\mathcal{T}_p^{i_1, i_{1^\prime}}$ = $\mathcal{T}_p^{i_{1^\prime}, i_1} = \mathcal{T}_p^{i_1^*, i_{1^\prime}^*}$ = $\mathcal{T}_p^{i_{1^\prime}^*, i_1^*}$ and together represent 36\% of the total contribution to the pressure transport.  The pressure transport transitions between a source and sink along the wake shear layer for $x/D \le 8$.  A similar distribution is obtained from the self interactions of the shedding frequency modes $\mathcal{T}_p^{i_1, i_1} = \mathcal{T}_p^{i^*_1, i^*_1}$, which account for 25\% of the total transport.   Similarly, the self interaction of the $\mathcal{T}_p^{i_{1^\prime}, i_{1^\prime}} = \mathcal{T}_p^{i^*_{1^\prime}, i^*_{1^\prime}}$ account for an additional 23\%.  Overall, the interactions of modes related to the shedding frequency contribute 84\% of the total transport indicating that the pressure fluctuations related to the shedding frequency transport CKE in the wake.   The interactions of the second and third harmonic contribute an additional 12\%, while the large changes of the spatial distribution are concentrated inside the wake shear layer. 

The contributions with shedding frequency mode pairs, all have a negative contributions immediately behind the cylinder where there is a negative correlation between the velocity fluctuations and the pressure fluctuations attributed to the shedding frequency DMD modes.  At $1 \le x/D \le 4$ theses modes become positive and slowly asymptote to near zero.  On the other hand, the contribution related to the second harmonic of the shedding frequency behaves opposite in the near wake and sharply increases to a maximum around $x/D=2$.  The peak and the second harmonic mode pair contribution dominate the total transport in the wake after $x/D=3$.  While the $i_1$ pairs (and their complex conjugates) affect the transport the most in the near wake, the far wake pressure transport is dominated by the coherent motions relate to the second harmonic, $i_2$.  

\begin{figure}   
   \begin{center}
       \includegraphics[width=\textwidth]{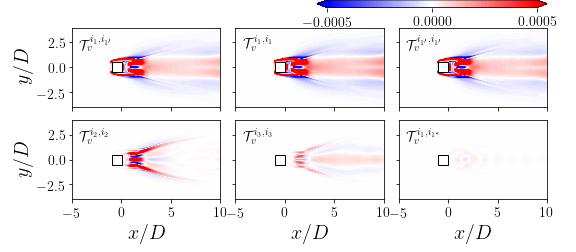}
       \caption{\label{fig:tv_top} Contours of six largest scale-specific contributions of the CKE transport via viscosity.}
    \end{center}
\end{figure}

The total CKE transport via viscosity $\mathcal{T}_v$ is nearly an order of magnitude smaller than CKE pressure transport $\mathcal{T}_p$ and turbulence transport $\mathcal{T}_t$.  Figure \ref{fig:tv_top} shows the six largest scale-specific contributions, $\mathcal{T}_v^{l,m}$, to the total CKE viscous transport.  Similar to the CKE pressure transport, the shedding frequency modes have the largest impact on the transport.  In fact, the three largest contributors (including the complex conjugates) $\mathcal{T}_v^{i_1, i_{1^\prime}}$, $\mathcal{T}_v^{i_1, i_1}$, and $\mathcal{T}_v^{i_1{1^\prime}, i_{1^\prime}}$ account of 82\% of all the CKE viscous transport.  All three have similar spatial distributions that are mainly non-zeros in the wake of the cylinder within the shear layer.  There is also high transport around the formation of the boundary layer on the cylinder.  The relative transport quickly decreases by $x/D=3$.  In the far wake, each scale-specific contribution is positive behind the cylinder and negative outside the wake.  The next two largest contributions are from modes associated with the second and third integer multiple and account for 10\% of the contribution.  Similar to the scale-specific CKE pressure transport, these spatial distributions differ significantly from the shedding frequency transport.  The ($i_2,i_2$) and ($i_3,i_3$) modes are active in the near wake and along the expanding shear layer, however, are not relatively large in the far wake.  The sixth largest scale-specific CKE viscous transport term is the interaction of the shedding frequency and its complex conjugate, $\mathcal{T}_v^{i_1, i^*_1}=\mathcal{T}_v^{i^*_1, i_1}$. While this only accounts for a 1\% contribution, it signifies that there is an interaction of forward and backwards frequencies modes.   

\begin{figure}
   \begin{center}
       \includegraphics[width=\textwidth]{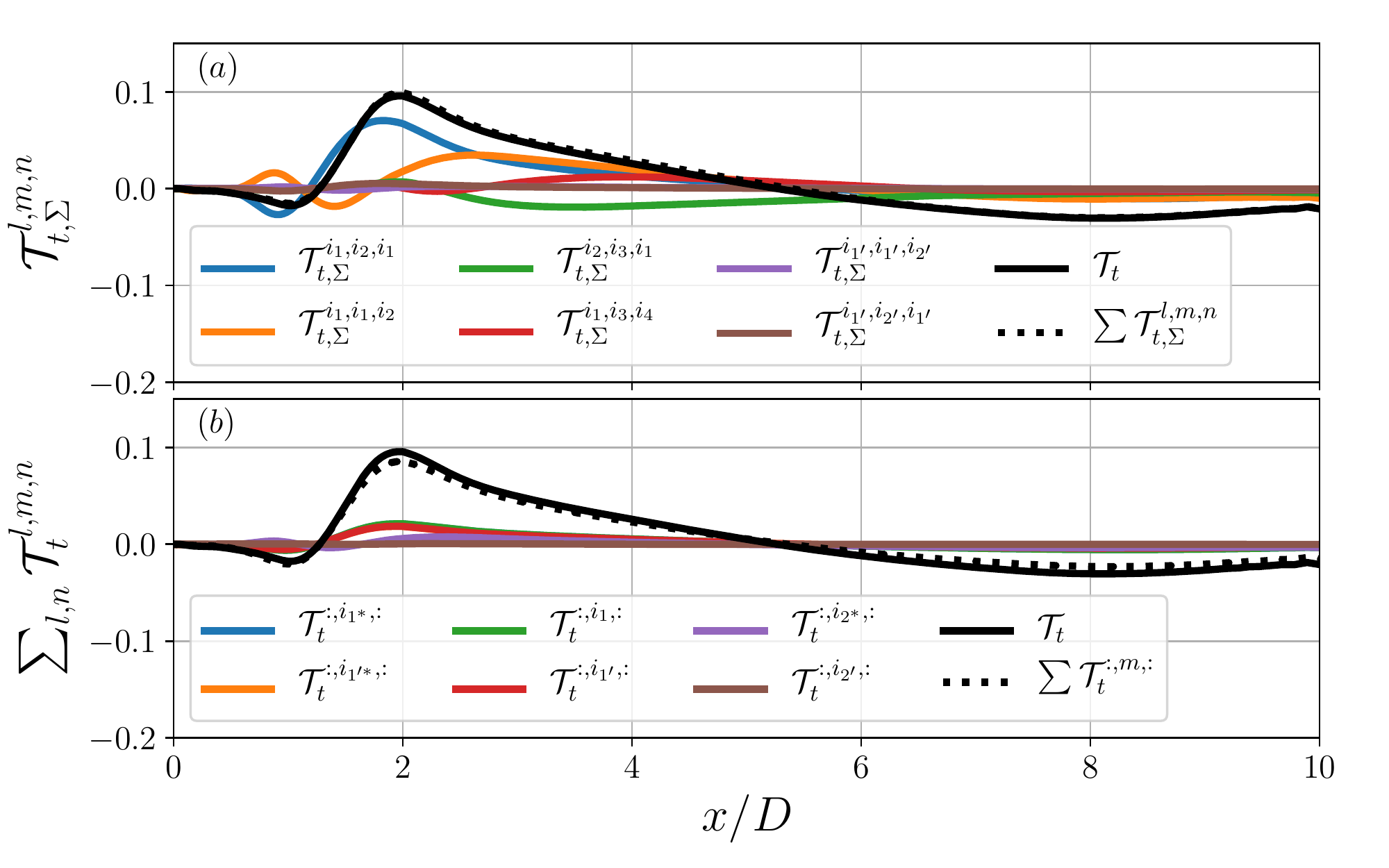}
       \caption{\label{fig:tt_centerline} Centerline, $y/D=0$, profile of the five largest scale-specific contributions of the CKE transport via turbulence. The total CKE turbulence transport $\mathcal{T}_t$ and sum of the five largest contributions $\sum \mathcal{T}_t$. (a) The sum of all contributions of $\mathcal{T}_t^{l,m,n}$ of the same class and (b) sum over all $l,n$ modes.}
    \end{center}
\end{figure}

The centerline profiles of the scale-specific turbulent transport $\mathcal{T}_t^{l,m,n}$ are shown in Fig. \ref{fig:tt_centerline}(a).  Similar to the classes of spatial distributions of scale-specific inter-scale transfer shown in Fig. \ref{fig:ptijk_mode2}, the permutations of triads have the same spatial distribution by different total magnitudes for scale-specific turbulent transport.   The five largest classes of triadic  contributions to the total CKE turbulence transport are shown in Fig. \ref{fig:tt_centerline}(a). The sum of all permutations over the triads in the class are obtained as $\mathcal{T}_{t,\Sigma}^{l,m,n} = \sum_{l,m,n \in C} \mathcal{T}_t^{l,m,n}$, where $C$ is the set of triads in the class.  The relative magnitude of each of the largest summations is comparable to the magnitude of the scale-specific CKE  pressure transport which indicates that both transport terms are mutually dominant in the total CKE balance.  The total CKE turbulence transport is also shown in Fig. \ref{fig:tt_centerline}(a).  The sum of the five largest classes of scale-specific CKE turbulence transport accounts for 99\% of the total. For all scale-specific distributions, the transport is highest along the centerline in the near wake and gradually decreases to a fraction of the peak in the far wake.  This is due to the expansion of the wake where there are only small fluctuations along the centerline in the far wake.   

Figure \ref{fig:tt_centerline}(b) presents an alternative summation of triad of the scale-specific turbulence transport through the profile of the turbulent transport along the centerline.  These centerline profiles are obtained by selecting the five largest dyadic pairs ($l,m$) of scale-specific CKE pressure transport in Fig. \ref{fig:tp_top} and summing over all the $l$ and $n$ modes in the triad.  This is used to interpret the role of scale-specific CKE turbulence transport on the scale-specific CKE budget in Eqn. \ref{eqn:ss_cke} as well as its relationship compared to the scale-specific CKE pressure transport.  From the profiles, we see that none of the profiles along the centerline have the same magnitude as the largest scale-specific pressure transport.  However, the summation of the five largest contributions combined to about 98\% of the total CKE turbulence transport. This presents further evidence that scale-specific CKE components are amplified (or attenuated) by the transfer of CKE via the triadic interactions.  

\begin{figure}   
   \begin{center}
       \includegraphics[width=\textwidth]{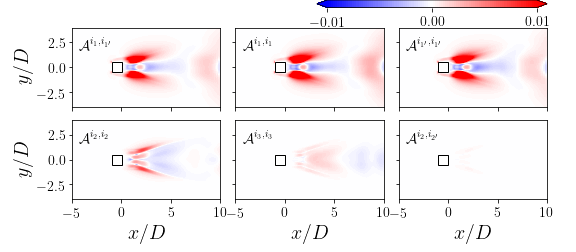}
       \caption{\label{fig:udk_top} Contours of six largest scale-specific contributions of the mean CKE advection.}
    \end{center}
\end{figure}

The mean CKE advection $\mathcal{A}$ has a significant contribution to the CKE budget in the wake of the cylinder with a magnitude on the same order as the production and CKE pressure transport. Scale-specific CKE advection distributions are shown in Fig. \ref{fig:udk_top}.  Again, we see that the shedding frequency mode pairs have the dominant effects on the total mean advection.  In the near wake, the effects of the shedding frequency pairs create similar distributions that alternate between source and sink of the CKE budget along the centerline and are strong along the shear layers of the wake.  Similar to the transport distributions for the shedding frequency pair, the relatively high contributions decrease near $x/D=5$ as the near wake transitions to the far wake.  The higher (second and third) integer multiples of the shedding frequency modes pairs have a similar contribution to the mean CKE advection compared the corresponding scale-specific CKE pressure transport.  The behaviour of these contributions to the total mean CKE advection are relatively similar to the shedding frequency, which is dominant in shear layers in the near wake.  Overall, the largest contributions are relatively similar along the centerline where the influence of the mean CKE advection is relatively low.  A major factor along the centerline is that the mode pairs of the shedding frequency and its higher harmonic frequencies are relatively similar through the downstream extent of the wake.  The sum of the five largest mode pair compares well to the total mean CKE advection indicating that these modes contain most of the information.  
 
\begin{figure}   
   \begin{center}
       \includegraphics[width=\textwidth]{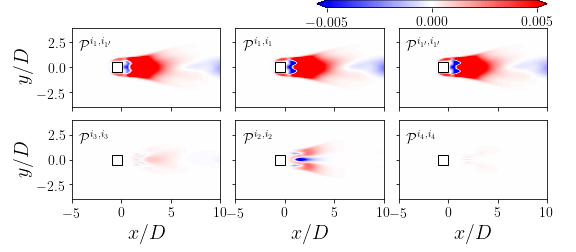}
       \caption{\label{fig:pr_top} Contours of six largest scale-specific contributions of the CKE production.}
    \end{center}
\end{figure}

The CKE production generally provides the energy from the mean flow to the largest scales in the flow.  With the developed methodology, contributions of the kinetic energy transfer from the mean flow can be identified with specific scales and coherent structures in the flow.   Figure \ref{fig:pr_top} shows the six largest mode pairs of the scale-specific CKE production.  Similar, the other dyadic scale-specific CKE terms, the most dominant pairs are the shedding frequency modes.  From the spatial distribution in comparison to the total CKE production available in Appendix \ref{sec:budg}, we can see that the shedding frequencies match the total distribution.  This further indicates that the shedding frequencies dominates the production of CKE.  Furthermore, these mode pairs account for 98\% of the total CKE production.  This is a clear indication that the turbulence energy is produced by the mean flow and enter directly into the modes related to the shedding frequency first.  From an energy cascade perspective, the energy is then mainly redistributed to the largest scales via the inter-scale transfer.  The other scales in the wake that have been shown to transport CKE must receive energy from nonlinear turbulent interactions.  The production provides clear evidence that the energy transfer mechanism to sustain the von K\'{a}rm\'{a}n vortices is due to CKE production from the mean flow directly directly into these large scales.  Relatively little energy is based from the mean flow into higher harmonic modes of the shedding frequency. The next three largest contributed from the third, second, and fourth harmonics of the shedding frequency.  The spatial extent of these shows major differences compared to the large shedding frequency production.  The second harmonic shows clear inverse production back to the mean flow along the centerline, but positive production along the shear layer.  This suggests that energy is continuously redistributed along in wake between the MKE and CKE.   

\begin{figure}   
   \begin{center}
       \includegraphics[width=\textwidth]{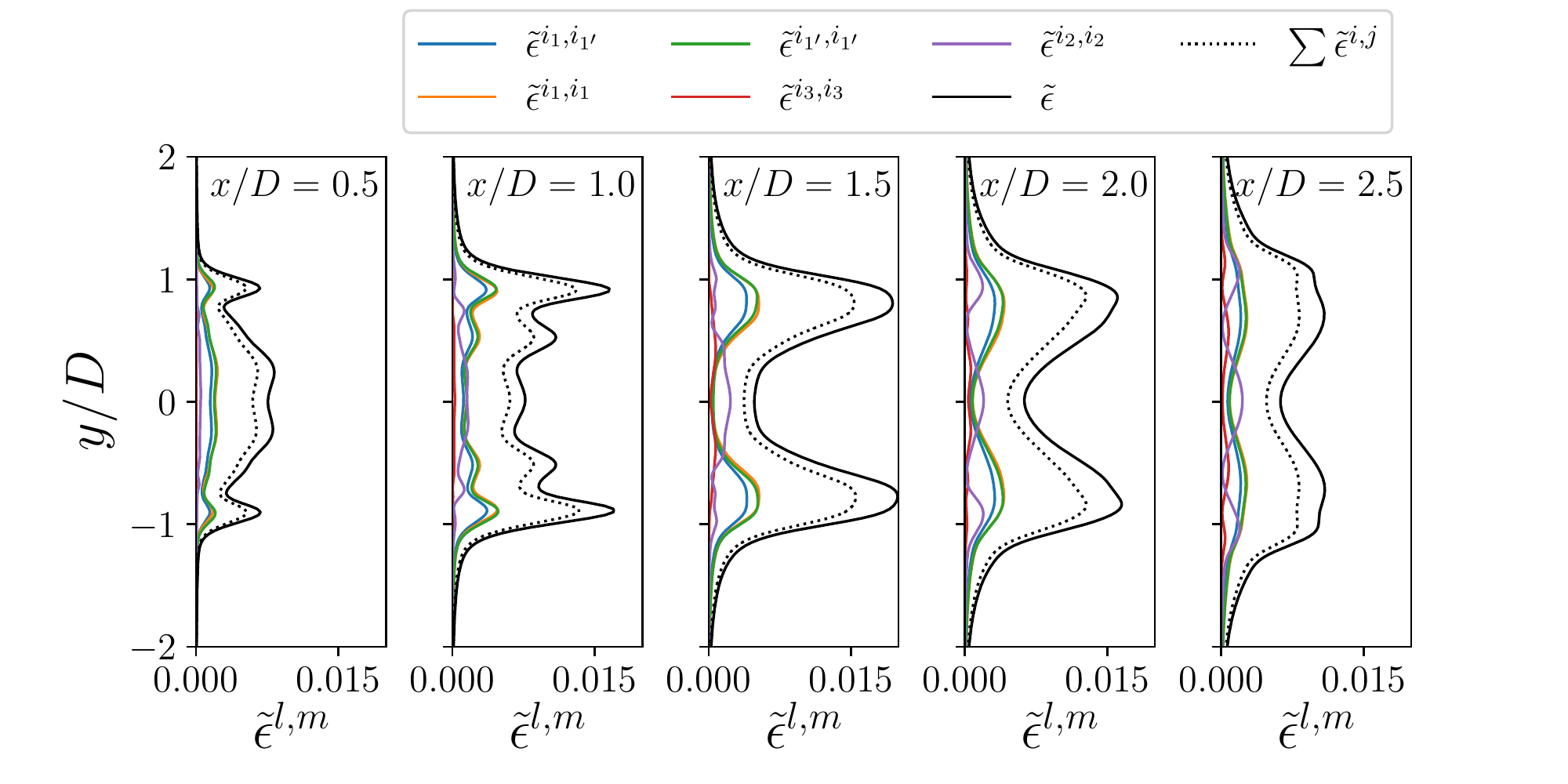}
       \caption{\label{fig:eps_profile} Profiles of five largest scale-specific CKE dissipation terms at $x/D = 0.5, 1, 1.5, 2$, and $2.5$. The total CKE dissipation $\tilde{\epsilon}$ and sum of the five largest contributions $\sum \tilde{\epsilon}$}
    \end{center}
\end{figure}

The scale-specific contributions of dyadic interactions of the CKE dissipation are shown in Fig. \ref{fig:eps_profile}.  The largest contributions are from the shedding frequency mode pairs. These scale-specific contributions have similar profiles, where it is maximum near the shear layer in the near wake and continues downstream along the shear layer. This contributions from CKE dissipation from modes related to the largest coherent structures in the flow present an interesting conclusion to how kinetic energy is dissipated from the wake.  First, we consider that this particular case is at a very low Reynolds number where the turbulence cascade has not developed.  This can readily be seen from the spectra in the wake shown in Fig. \ref{fig:spectra}. This indicates that there is not a clear mechanism in this flow that would allow viscous dissipation to only exist at the smallest scales.   In fact, at this Reynolds number, one would expect the energy and dissipation spectra to overlap each other.  Second, the turbulence cascade theory is based on length scales rather than time scales which separate the DMD modes created within this methodology.  While the Taylor's hypothesis is applicable in large area of the wake of the cylinder, it would not be applicable in the near wake. Furthermore, the convection velocity that would be employed in Taylor's hypothesis to realize length scales from time scale is not trivial for a DMD mode because each has a spatial distribution of velocity.   Due to these two points, we expect high dissipation associated with the dominant coherent structures identified through the DMD modes.   

Profiles of the scale-specific CKE dissipation and the largest mode pair contributions in Fig. \ref{fig:eps_profile} are shown at locations $x/D = 0.5, 1, 1.5, 2$, and $2.5$. The three pairs related to the shedding frequency all have similar profiles and dominant the overall profile of the CKE dissipation.   At the $x/D = 0.5$ position, the contributions from the largest three contributions are an order of magnitude larger than other mode pairs.  However, by $x/D=1$, the second and third harmonics for the shedding frequency begin to increase. The peaks of the second harmonic in the shear layer occur at local minima of the shedding frequency modes.  Along the centerline, the contribution of second harmonic mode pair becomes larger than the shedding frequency. This continues downstream at the other locations.  Furthermore, at downstream locations $x/D=2$ and $2.5$ the local maxima of the second harmonic mode pair in the shear layer increases and has a significant effect on the distribution of dissipation in the wake shear layer.  

\begin{figure}    
   \begin{center}
       \includegraphics[width=\textwidth]{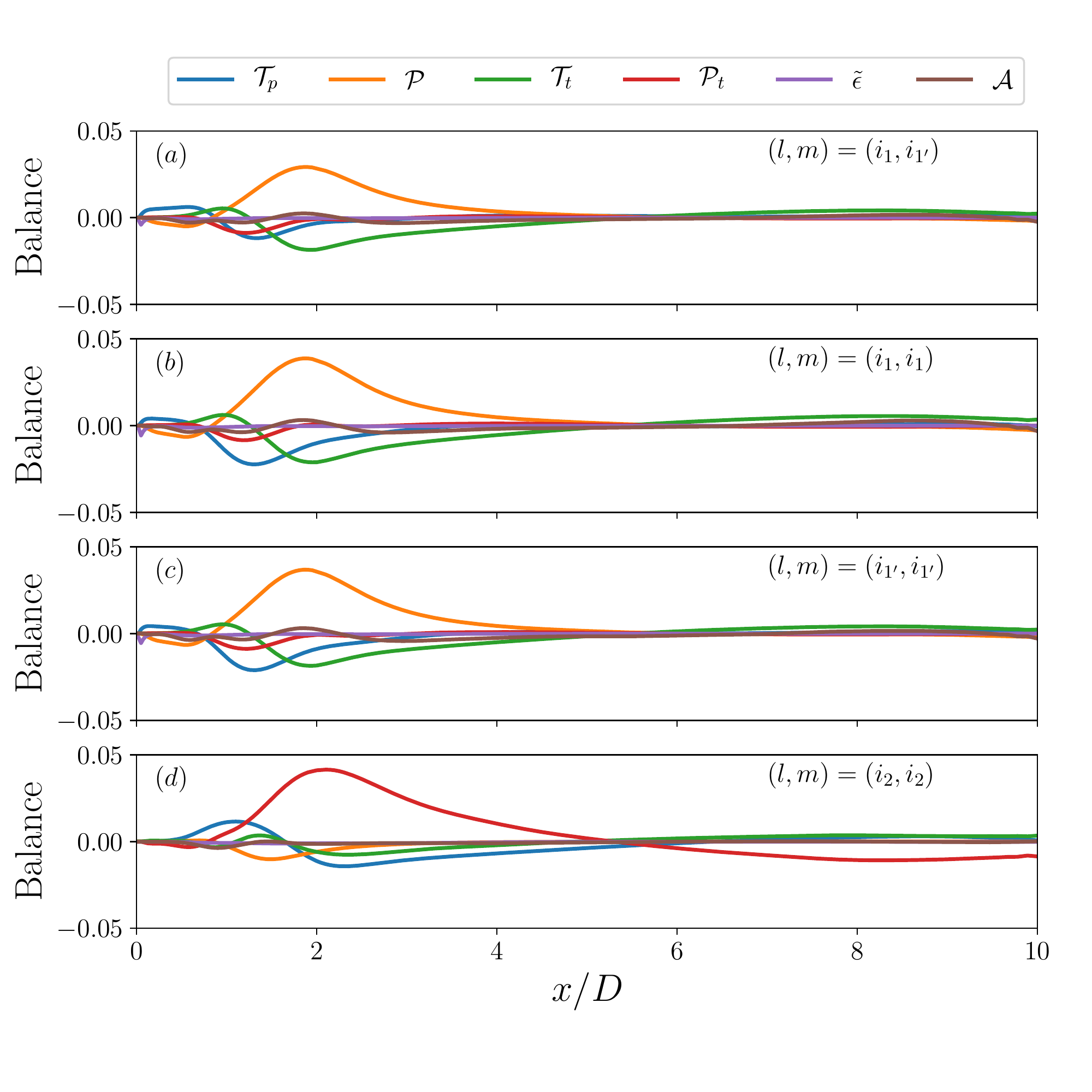}
       \caption{\label{fig:largest} Scale-specific CKE balance for the four largest scale contributions: (a) $(l,m) = (i_1,i^\prime_1)$, (b) $(l,m) = (i_1,i_1)$, (c) $(l,m) = (i^\prime_1,i^\prime_1)$, and (d) $(l,m) = (i_2,i_2)$. }
    \end{center}
\end{figure}

Figure \ref{fig:largest} shows the scale-specific CKE balance based on Eqn. (\ref{eqn:short_cke_ss}) for the four largest mode pair contributions to the scale-specific CKE along the centerline.  For the total CKE balance, see Appendix \ref{sec:budg}.  The largest contribution is from $(l,m) = (i_1,i^\prime_1)$, shown in Fig. \ref{fig:largest})(a).  The CKE production $\mathcal{P}$ is positive indicating that kinetic energy is transferred from the mean flow to scales associated with vortex shedding.  This is mainly balanced by the pressure and turbulence CKE transport at this scale.  The inter-scale transfer is negative indicated that scale-specific CKE is removed from this vortex shedding scale and transferred to other scales.  This is in agreement with analysis of the inter-scale transfer, which showed that transfers with $m=i_1$ mainly are negative indicating transfer to other scales.   Figure \ref{fig:largest}(b) and \ref{fig:largest}(c) show contributions from $(l,m) = (i_1,i_1)$ and $(l,m) = (i^\prime_1,i^\prime_1)$, which describe the evolution of CKE along the centerline for other scale interactions associated with the vortex shedding frequency.  All three scale-specific CKE balances behave the same.   On the other hand, Fig. \ref{fig:largest}(d) shows contributions from the  $(l,m) = (i_2,i_2)$ and has a very different balance of energy.  For this scale, associated a multiple of the vortex shedding, the main gain of CKE is through the inter-scale transfer from other scales. The contributions of all the modes associated with vortex shedding modes $\pm i_1, i^\prime_1$ mainly contribute to the large positive inter-scale transfer as well as other scales as shown in Fig. \ref{fig:ptijk_interact}.  This is balanced by the pressure and turbulence transport. Furthermore, the CKE production from this scale is negative indicating that energy transfers from this scale to the mean flow.  Overall, the largest scale-specific balances show how energy is transferred between scales.

\subsection{Unsteady inflow effects on energy transfer}
In this section, we introduce two additional test cases pertaining to the flow over a square cylinder with variable inflow condition.  This is performed to highlight the ability of the developed energy transport methodology to identify, characterize, and compare the modes of the energy transfer. The inflow is given as $U_{in} = U_\infty(1 + \frac{1}{10} \sin (2 \pi f_f t))$, where $f_f$ is the inflow imposed frequency.  The two cases are chosen based on the non-dimensional frequency $\mathrm{St}_f = f_f D/U_\infty$: (1) a multiple and resonance frequency of the shedding frequency $\mathrm{St}_f = 8.0$ and (2) a frequency that is not amplified in the steady flow case $\mathrm{St}_f = 2.5$.  The former case highlights energy transfer behaviour where energy is added at dominant flow frequency, while the latter demonstrates energy transfer behaviour where new energy modes are created by external forcing.  Each case modifies the spectral energy distribution, in particular, the modes related to the forcing frequency $f_f$ is identified and amplified compared to the steady case.

Each case is simulated in the same manner as the steady case and flow field snapshots are collected after initial transients are removed from the flow field.  Snapshot matrices with $M=4000$ are created with a $\Delta t = 0.02$, the same time step used in the steady simulation.  Mode decomposition and sparse sampling are performed to obtain a reduced set of DMD modes.  The number of DMD modes selected that can be accurately reconstruct the flow field and capture the TKE as discussed for the steady case is $R=14$ and $R=34$ for the $\mathrm{St}_f = 8$ and $\mathrm{St}_f = 2.5$, respectively.  For both cases, the number of modes selected represent an error in velocity reconstruction and TKE to be $\epsilon_u < 0.1$ and $\epsilon_k < 0.02$, respectively.      

\begin{figure}   
   \begin{center}
       \includegraphics[width=\textwidth]{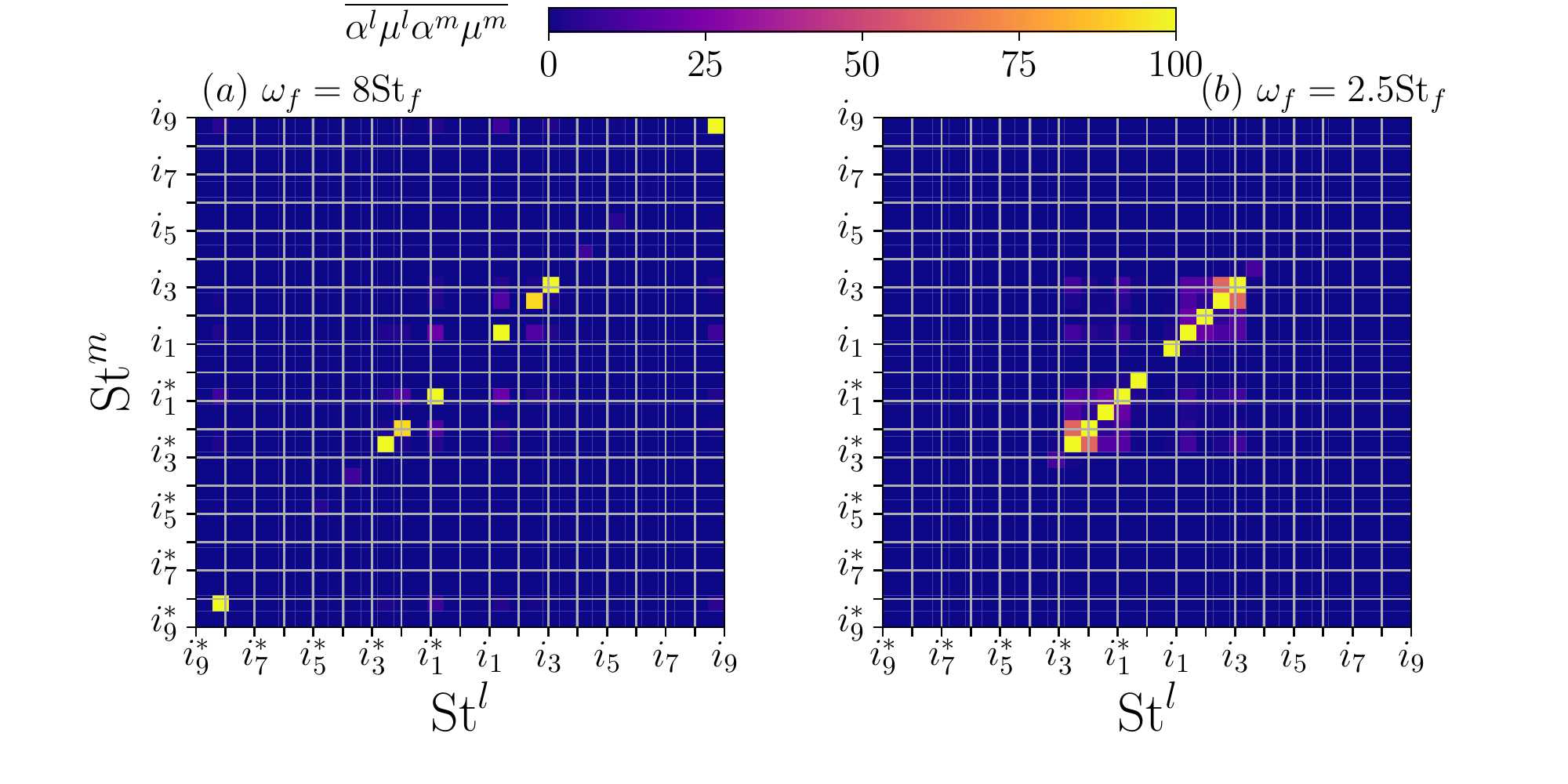}
       \caption{\label{fig:forced_corr}The dyadic correlation of the DMD modes for (a) $\mathrm{St}_f = 8$ and (b) $\mathrm{St}_f = 2.5$. }
    \end{center}
\end{figure}

The correlations of the modes as a function of the $l$ and $m$ frequencies, $\overline{\alpha^l\mu^l\alpha^m\mu^m}$, are shown in Fig. \ref{fig:forced_corr}(a) and (b) for cases $\mathrm{St}_f = 8$ and $\mathrm{St}_f = 2.5$, respectively.  The dyadic correlations of the $\mathrm{St}_f = 8$ are similar to the steady inflow case.  The maximum correlation occurs along the diagonal where $l = m$.  For example, a highly correlated dyadic is a self-conjugate correlation: $(l,m)=(i_1, i^*_1)$. The correlations are highest near the shedding frequency and diminished as the frequency increases.  However, another peak correlation is observed that does not occur in the steady inflow case at the forcing frequency of $\mathrm{St}_f = 8i_1 = i_8$.  Additionally, a moderate correlation at $(l,m) = (\pm i_1, \pm i_8)$ suggests that there is an exchange of energy directly between the forcing frequency and the shedding frequency.  Correlation to other harmonics of the shedding frequency occur at significantly lower levels.   

Dyadic correlations for the $\mathrm{St}_f = 2.5$ exhibit some similarities to the steady inflow case, but have a significantly large range of frequencies that are highly correlated.  Unlike the $\mathrm{St}_f = 8$ case, the inflow frequency for this case is not a harmonic of the shedding frequency and amplifies modes that are not related to shedding.  Not only are there self-conjugate correlations between the shedding frequency and its harmonic multiples, but frequencies related to multiples of the forcing frequency as well.  There are also high correlations between the shedding frequency and forcing frequency, all of which occur off the diagonal of $l=m$.  Unlike the $\mathrm{St}_f=8$ case where only the shedding and forcing modes are correlated, this case exhibits high correlations between harmonics of both frequencies. This is a main contributor to why this case requires more than twice the number of modes to adequately capture the velocity dynamics and TKE.  

\begin{figure}   
   \begin{center}
       \includegraphics[width=\textwidth]{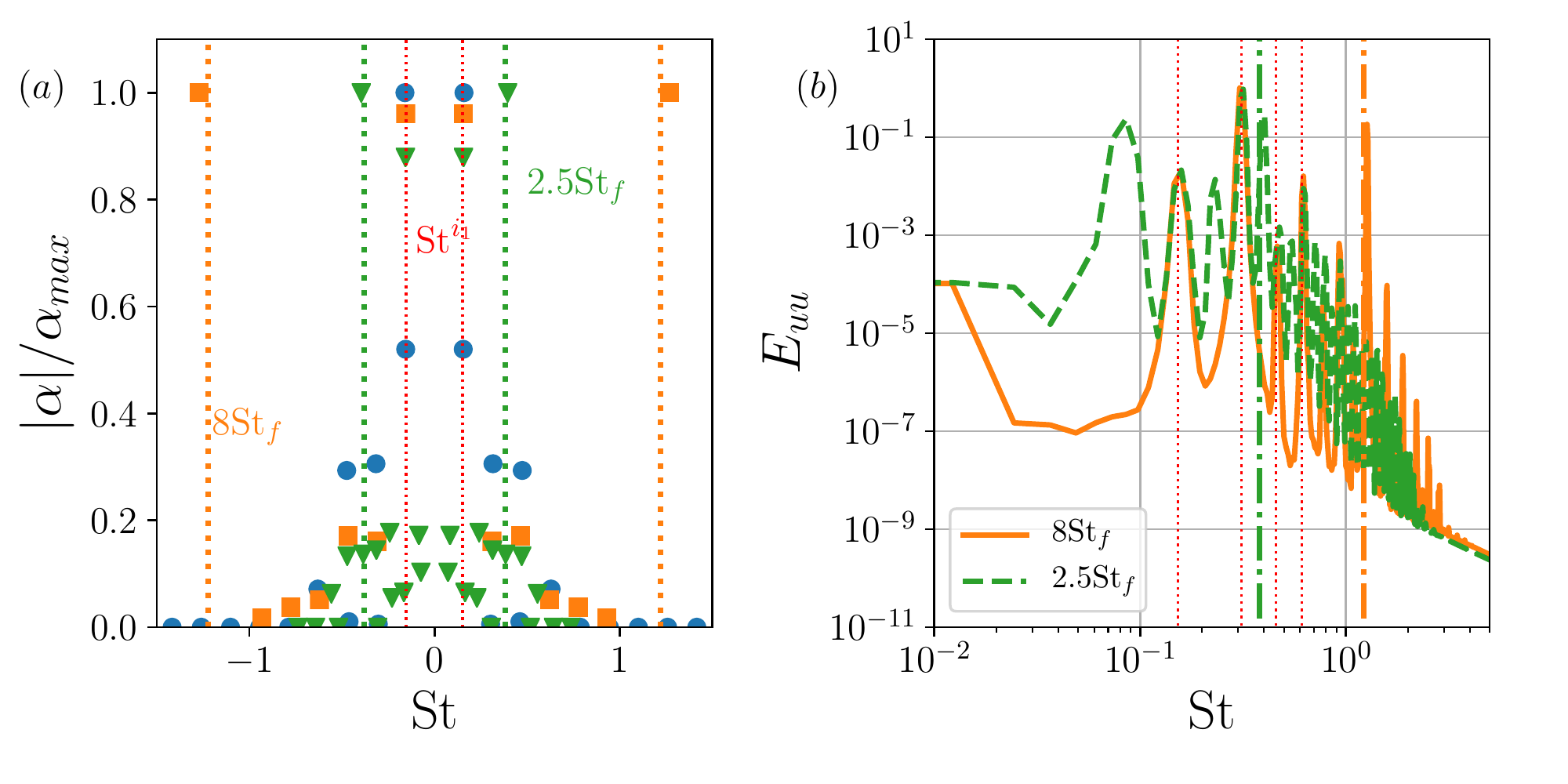}
       \caption{\label{fig:forced_amp} (a) The DMD amplitudes normalize by the maximum amplitude $\alpha/\alpha_{max}$ of the steady inflow case (circle), $\mathrm{St}_f=8$ case (square), and $\mathrm{St}_f = 2.5$ (triangle) and (b) the streamwise energy spectra $E_{uu}$ of the $\mathrm{St}_f=8$ case and $\mathrm{St}_f = 2.5$ at $x/D=9$ along the centerline.}
    \end{center}
\end{figure}

Figure \ref{fig:forced_amp}(a) shows the DMD amplitudes normalized by the maximum amplitude of the steady inflow, $\mathrm{St}_f=8$, and $\mathrm{St}_f = 2.5$ cases.  Each unsteady inflow case exhibits a different spectrum, where the largest amplitude is associated with the forcing frequency, while the second largest amplitude is the shedding frequency.  Additional high amplitudes quickly decay for the $\mathrm{St}_f=8$ case.  The decay is similar to the steady inflow case, where each amplitude is associated with harmonics of the shedding frequency. Additional high energy modes for the $\mathrm{St}_f = 2.5$ case are observed at frequencies smaller than the shedding frequency and larger than the forcing frequency.  The distribution of energy of DMD modes is confirmed by the streamwise velocity energy spectra at $x/D=9$ for the unsteady inflow cases. Both spectra are populated by energy in more frequency modes compared to the steady inflow case in Fig. \ref{fig:spectra}. The spectra confirm that the highest energy is located in modes associated with the forcing frequency, while the shedding frequency has the second highest energy.   

The spectra of the scale-specific inter-scale transfer over all $m$ modes $\sum_{m=0}^R \mathcal{P}_t^{l,m,n}$ is shown for the $\mathrm{St}_f=8$ case at $x/D=1$ for $y/D=0, 0.25, 0.5$ and $1$ locations.  Similar to Fig. \ref{fig:ptijk_interact_all}, the contributions of all $m$ modes are summed at each location to identify the magnitude and direction of inter-scale transfer to the $l,n$ modes.  The triadic interactions are anti-symmetric across the $\mathrm{St}^l = \mathrm{St}^n$ line.  At the $y/D=0$ and $0.25$ locations, overall energy is transferred from low frequency modes associated with the shedding frequency and to modes associated with higher frequencies.  Similar to the steady case in Fig. \ref{fig:ptijk_interact_all}(a), the spectra indicates that energy primarily transfer from lower frequency scales to higher frequency scales in the near wake.  At these location in the wake, the forcing frequency mode does not have a large impact on the inter-scale transfer of CKE.  Further from the centerline at $y/D=0.5$, which is located near the shear layer, the inter-scale transfer behaves similarly to the locations in the wake.  However, outside of the wake at $y/D=1$, the inter-scale transfer is only among a few low frequency modes. Similar to the steady case at this location, the CKE is transferred to those low frequency modes.  While the forced case behaves similar to the steady case, the spectra reveals that interactions occur among more scales compared the steady case. There is high levels of transfer between high frequencies up to scales associated with 4 times the shedding frequency. These differences suggest that the additional energy forced into the flow effects the transfer.   However, at this location in the wake, the effects of the unsteady inflow are not clear because the near wake blocks most of the energy in the incoming flow.  

\begin{figure}
   \begin{center}
       \includegraphics[width=\textwidth]{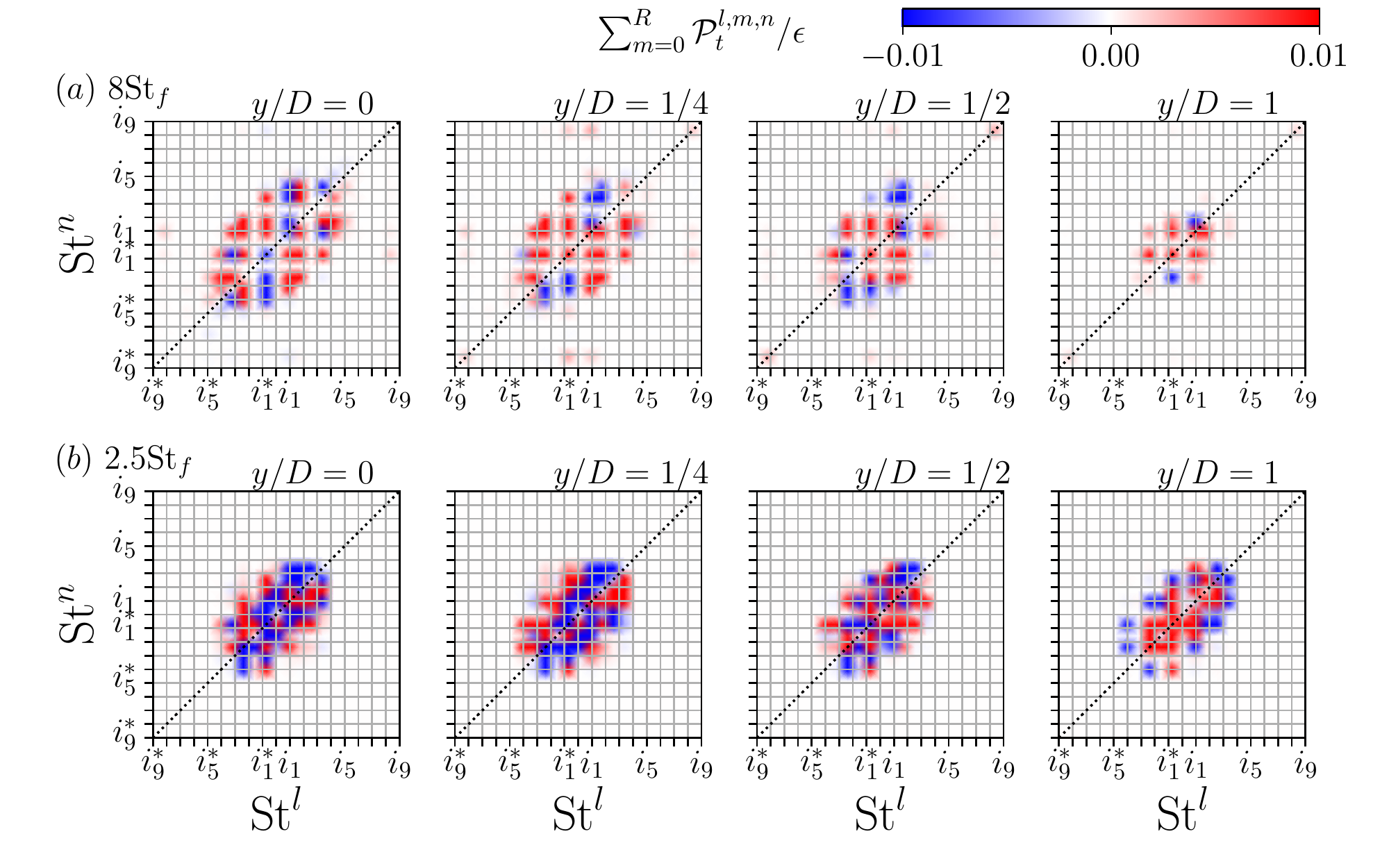}
       \caption{\label{fig:ptijk_int_force1} Spectra of the sum over $m$ scale-specific inter-scale transfer $\sum_{m=0}^R \mathcal{P}_t^{l,m,n}$ normalized by the total turbulence dissipation at $y/D=0, 0.25, 0.5$, and $1$ in the near wake at $x/D=1$ for the (a) $\mathrm{St}_f = 8$ and (b) $\mathrm{St}_f = 2.5$ cases.}
   \end{center}
\end{figure}

The spectra of the scale-specific inter-scale transfer of the $\mathrm{St}_f = 2.5$ case are significantly different that the other two cases present as shown in Fig. \ref{fig:ptijk_int_force1}(b). The forcing frequency at the inflow amplifies several scales unrelated to the shedding frequency.  While the other unsteady case, did not have a large an effect of the incoming flow frequency in the near wake, this case clearly shows CKE transfer away from the forcing frequency as well as other low frequency modes. However, there is also positive transfer to mode pairs (and conjugates) of ($i_1$ and $f_f$), which indicates that the modes of the shedding frequency and forcing frequency interact and the CKE associated with that interaction increases.  The three locations in the wake ($y/D=0,0.25,$ and $0.5$) exhibit similar spectra.  In the location outside the wake, $y/D=1$, the trends of CKE transferring to low frequencies is similar to the other cases.   

\begin{figure}
   \begin{center}
       \includegraphics[width=\textwidth]{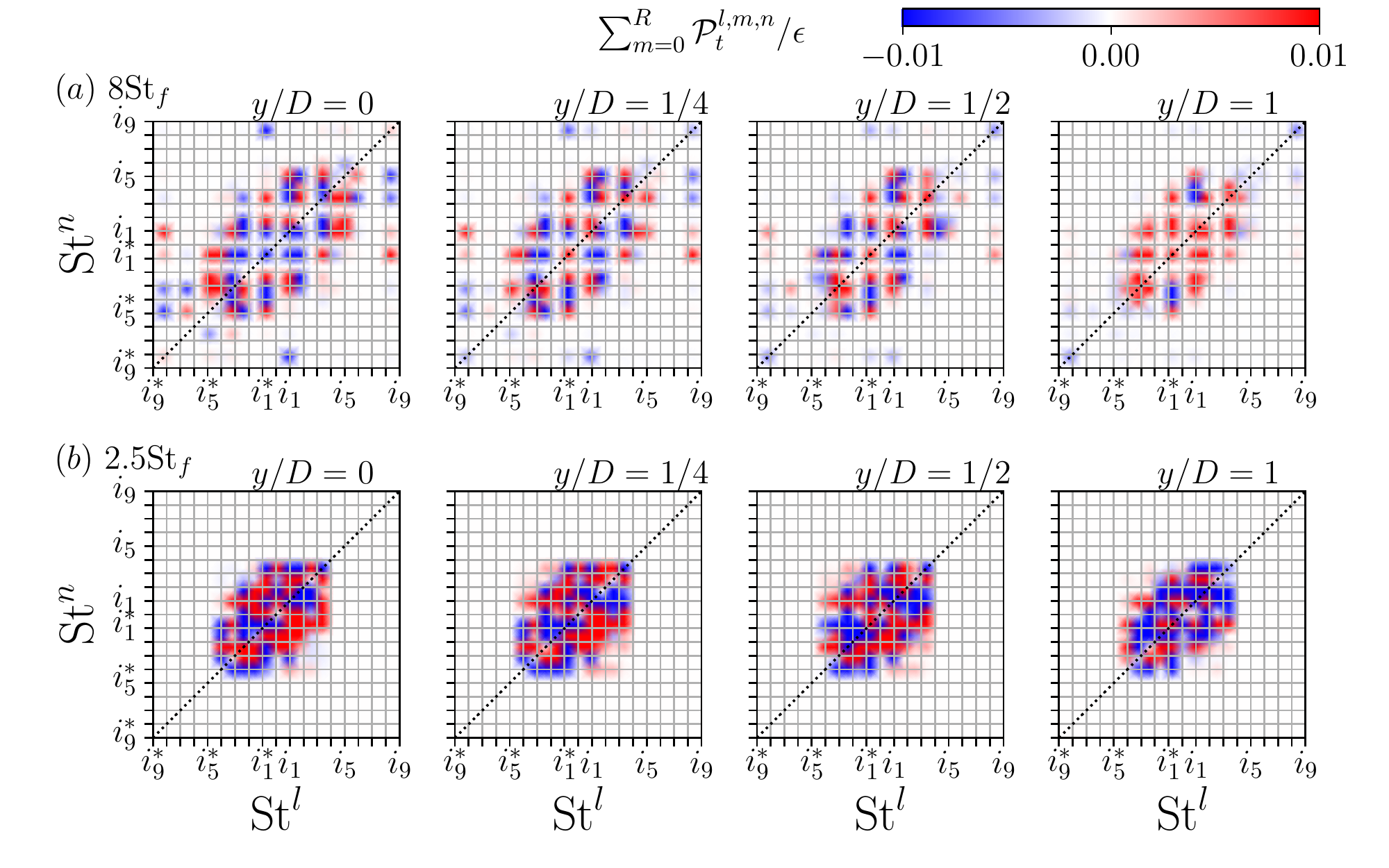}
       \caption{\label{fig:ptijk_int_force8} Spectra of the sum over $m$ scale-specific inter-scale transfer $\sum_{m=0}^R \mathcal{P}_t^{l,m,n}$ normalized by the total turbulence dissipation at $y/D=0, 0.25, 0.5$, and $1$ in the near wake at $x/D=8$ for the (a) $\mathrm{St}_f = 8$ and (b) $\mathrm{St}_f = 2.5$ cases.}
   \end{center}
\end{figure}

At a further downstream location in the far wake, $x/D=8$, the spectra of the scale-specific inter-scale transfer shows a large impact of the forcing frequency for the $\mathrm{St}_f=8$ shown in Fig. \ref{fig:ptijk_int_force8}(a).  The inter-scale transfer is associated energy going from modes associated with low frequency and towards high frequency modes. Additionally, there are broad interactions between the forcing frequency and many other scales associated with integer multiples of the shedding frequency. The transfer of energy is typically from these interactions.  On the other hand, interactions between the forcing frequency and the shedding frequency are positive indicating that scale-specific CKE of those modes is increased through those interactions.  This behaviour continues at larger radial location.  
The downstream spectra of the $\mathrm{St}_f=2.5$ case at $x/D=8$ shows interactions that exhibit similar behaviour compared to the $x/D=1$ location.   
Figure \ref{fig:ptijk_int_force8}(b) shows CKE is transferred to a broad range of modes  throughout the wake, which suggests that the inter-scale transfer has become more homogeneous in the far wake.

\section{Conclusions}\label{sec:conclusions}
The scale-specific CKE evolution equation developed in this work offers the opportunity to study the implications of interactions of the specific scales associated with coherent structures in a flow.  Various terms in the evolution equation are shown to reveal impacts of energy transfer and transport.  The former is quantified by both (1) turbulence production terms, which transfer energy among mean, coherent, and random scales and (2) inter-scale transfer, which appears in the scale-specific CKE equation and identifies how coherent scales transfer kinetic energy among themselves.  The scale-specific inter-scale transfer terms are based on triadic interactions of coherent scales and formulated in a production-like manner, where kinetic energy is transferred from the velocity gradient of one coherent scale to the CKE associated with two coherent scales.  The  scale-specific inter-scale transfer allows us to quantify kinetic energy transfer among scale.   This is particularly useful in multi-scale flows that exhibit inhomogeneity and anisotrophy due to the effects of the coherent structures.  On the other hand, energy transport terms are generally dyadic, or related to two scales, terms.  
In this work, we quantified coherent scales in the scale-specific CKE equations based on the dynamic mode decomposition of the spatio-temporal flow field.  This is employed because coherent scales in the flow are identified via their frequency and modes are non-orthogonal, which enable us to quantify the CKE of interactions between two modes.  

The scale-specific kinetic energy methodology was applied to a low Reynolds number wake flow of a square cylinder.  This flow produces dominant coherent structures in the near wake that persist into the far field of the wake.  The low Reynolds number was tested to ensure that there are relatively few scales present in the flow to simplify the assessment of the methodology.  Compressive sensing is employed to find an optimal number of modes compared to the reconstruction error and error compared to the TKE.  The number of modes chosen attempts to minimize the error in TKE to ensure that the random kinetic energy is negligible and enable the identification of all interactions from the largest to the smallest scales.  The selection of modes was shown to have little influence on the scale-specific CKE of the most energetic modes. While the metric may not be an acceptable for most flows, especially high Reynolds number flows with many scales, in the present case due to the low Reynolds number the number of energetic scales is low, but dominant features exist in the flow.  

The scale-specific CKE reveals that relatively few modes contain most of the coherent kinetic energy.  The largest contributions reside in the energy of the modes related to the shedding frequency followed by the first harmonic (twice the frequency) of the shedding frequency.  The composition of CKE is mainly focused on the energy in these two modes as well as energy present in the interactions of the two modes.  Similarly, the scale-specific transport and dissipation terms are also associated with the dominant modes of the flows, which confirms that vortex shedding transports the majority of the CKE.
The scale-specific coherent production shows that coherent kinetic energy is injected from the mean flow into the shedding frequency mode.   Triadic interactions appear in both turbulence transport and inter-scale transfer.  In the former, the effects of the transport via velocity fluctuations is of similar scale as the transport via pressure and are dominated by the fluctuations of the vortex shedding.  The latter terms show that there is a wide range of transfer between coherent structures.  In most locations in the wake, the CKE is transferred from modes associated with vortex shedding to higher frequency modes, which are associated with integer multiples of the shedding frequency. We also found the feedback of energy to lower frequency scales.   

The present work constructs a methodology to analyse the energy transfer between coherent structures in a flow.  Future work will consider the interactions in fully turbulent flows, where there is a broad range of scales and the energy containing scales are separated from the dissipation scales.  The range of modes selected will have to be carefully considered due to the large number of modes that can be created.   Overall, the present work shows promise in elucidating details of coherent structure behaviour with respect to energy transfer.

\begin{acknowledgements}
This work was supported by the University of Memphis.
\end{acknowledgements}
\section*{Declaration of Interests}
The authors report no conflict of interest.

\appendix
\section{Dynamic Mode Decomposition and Compressive Sensing}\label{sec:dmd}
In this section, we detail to the algorithm used for sparsity-promoting DMD~\citep{jovanovic2014sparsity}.  We define two matrices such that one is offset from the other by one time instance as follows:
\begin{align*}
        X= 
        \begin{bmatrix}
            \mid & \mid & ... & \mid\\
            x_{1} & x_{2} & ... & x_{M-1}\\
            \mid & \mid & ... & \mid
       \end{bmatrix}, \:\:\:
        X^{'}= 
        \begin{bmatrix}
            \mid & \mid & ... & \mid\\
            x_{2} & x_{3} & ... & x_{M}\\
            \mid & \mid & ... & \mid
       \end{bmatrix}
    \end{align*}
where, $X, X^\prime \in \mathcal{R}^{N \times M}$, $M$ is the number of degrees of freedom in a snapshot, and $M$ is the number of snapshots. Each snapshot $x_i$ is uniformly sampled in time separated by a time step $\Delta t$.  
We employ the DMD algorithm derived in \cite{schmid2010dynamic}. The SVD of $X$ is computed as:
\begin{equation}
X = U \Sigma V^{T}
\end{equation}
where $T$ denotes the transpose,  $U \in \mathcal{R}^{N \times S}$  are the left singular vectors, $\Sigma \in \mathcal{R}^{S \times S}$ is a diagonal matrix of the singular values, $V \in \mathcal{R}^{M \times S}$ are the right singular vectors, and $S$ is the rank of the reduced SVD.
A reduced linear operator $\tilde{A} \in \mathcal{R}^{S \times S}$ can be efficiently obtained by projecting $A$ with the left singular vectors $U$ as follows:
\begin{equation}
\tilde {A}= U^{*}A U = U^{*}X^{'}V \Sigma ^{-1}
\end{equation}
The matrix $\tilde{A}$ is the reduced mapping of the dynamical system.  Spectral information of $\tilde{A}$, which has been shown to be the same as $A$, is obtained through an eigen-decomposition of $\tilde{A}$:
\begin{equation}
\tilde{A}W=W\Lambda
\end{equation}
where columns of $W \in \mathcal{C}^{S \times S}$ are eigenvectors and $\Lambda \in \mathcal{C}^{S \times S}$ is a diagonal matrix containing the corresponding eigenvalues $\lambda = \lambda_r + \imath \lambda_{i}$.
One can obtain the more familiar complex frequency, $\imath \omega_r = \log(\lambda_r) / \Delta t$. The real part is the temporal frequency, and the imaginary part is an exponential growth rate of the dynamic mode.  The matrix of the spatial dynamic modes $\Phi \in \mathcal{C}^{M \times S}$ are recovered with the following:
\begin{equation}
\Phi = X^{'}V\Sigma ^{-1}W
\label{eqn:dmd_modes}
\end{equation} 

The optimal amplitudes $b$ can be solved through an $\mathrm{L}2$ minimization as the following:
\begin{equation}
  \underset{{\alpha}}{\mathrm{minimize}} \quad J(\alpha) = \|X-\Phi D_{\alpha} V_{and} \|^{2}_{2}, 
\end{equation}
where $V_{and} \in \mathcal{R}^{S\times R}$ is the Vandermode matrix of the eigenvalues $\Lambda$. The row of the $V_{and}$ associated with the $k$th eigenvalue is $\mu^k$.  
The calculation of the optimal amplitudes was simplified by ~\cite{jovanovic2014sparsity} to the following:
\begin{equation}
   J(\alpha) = \alpha^T P \alpha - q^T\alpha - \alpha^Tq + s,
\label{eqn:opt}
\end{equation}
where $P=(W^T W) \odot (V_{and} V_{and}^T)^*$, $q = (\mathrm{diag}(V_{and}V \Sigma W))^*$, and $s = \mathrm{trace}(\Sigma^T \Sigma)$.    A sparse solution is induced by including  the L1-norm of vector $\alpha$ to the optimization problem in Eqn. (\ref{eqn:opt}):
\begin{equation}
    \underset{{\alpha}}{\mathrm{minimize}} \quad J(\alpha) + \gamma \sum_{i=1}^{R} |\alpha_{i}|
    \label{eqn:sp_opt}
\end{equation}
where, $\gamma$ is a positive parameter that controls the sparse solution of the amplitudes vector $\alpha$.  The solution is obtained by solving the optimization using the alternating direct method of multiplies.

\section{Kinetic Energy Budgets}\label{sec:budg}
Here we focus on the budgets of the MKE and CKE components of the kinetic energy.  The MKE is obtained from temporal averaged statistics obtained from the flow field over $M=4000$ time steps, the size of the snapshot matrix and over $40000$ time steps.  The flow field statistics converge over $M$ time steps such that the MKE terms are the nearly identical both time ranges of the statistics.  The CKE components are obtained directly by summing over the scale-specific CKE terms from the $S=31, R=24$ case.    

\begin{figure}
   \begin{center}
       \includegraphics[width=\textwidth]{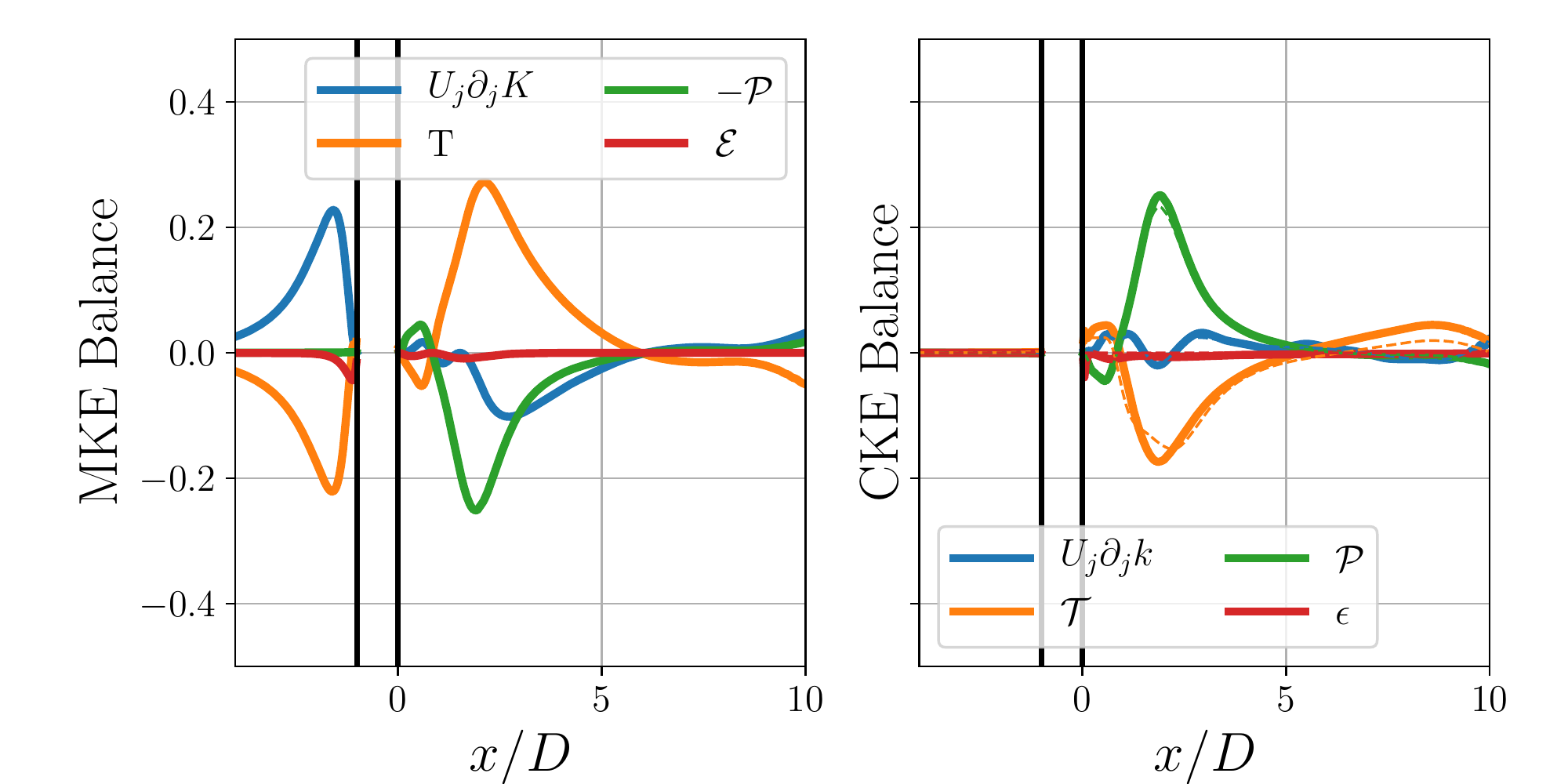}
       \caption{\label{fig:budget_centerline} (a) MKE and (b) CKE budget along the centerline, $y/D=0$ for the $S=31, R=24$ case. Dashed lines in (b) represent the TKE budget terms obtained from simulation statistics.}
    \end{center}
\end{figure}

Figure \ref{fig:budget_centerline}(a) and (b) shows the MKE and CKE budget along the centerline, $y/D=0$ of the wake where the maximum production is located.  The profiles along the centerline confirm that the terms balance each other, as expected.  Along the centerline we see that the mean MKE advection is balance by the transport upstream of the wake.  Immediately downstream of the cylinder the transport, mainly transport via mean pressure, is balanced by the production.  This is region where turbulent energy is being produced.  Further downwind, the mean MKE advection increases as the production decreased and is balanced by the transport.  The CKE budget reveals that the CKE produced along the centerline balance by transport, mainly by pressure forces.  Figure \ref{fig:budget_centerline}(b) also show the TKE budget obtained from statistics over $40000$ time steps. All TKE terms are nearly the same as the CKE terms.  Discrepancies in the profiles are attributed to the separation of coherent and random components.  The latter are not accounted for the CKE balance.  Furthermore, the TKE balance is similar to results obtained in \cite{saha2000vortex} at a similar Reynolds number.  

\begin{figure}
   \begin{center}
       \includegraphics[width=\textwidth]{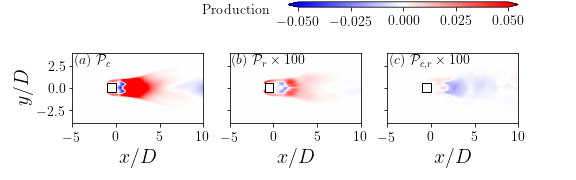}
       \caption{\label{fig:production} Contours of (a) Coherent production $\mathcal{P}_c$, (b) coherent-random production $\mathcal{P}_{cr}$, and (c) random production $\mathcal{P}_r$ for the $S=31, R=24$ case.}
    \end{center}
\end{figure}

Production connects the three equations of the kinetic energy budgets together and represent the redistribution of energy into the different components.  Figure \ref{fig:production}(a) shows the coherent  production rate $\mathcal{P}_c$, which appears in both the MKE and CKE budgets. This accounts for the production of coherent kinetic energy from the mean strain rates.  This production occurs in the wake and along the boundaries of the cylinder.  The production mainly appears directly behind cylinder in the near-field region of the wake.  In this region the velocity gradients and the Reynolds stresses are the highest.  It is mainly co-located where the CKE is distributed. The coherent production is not entirely positive.  There are two regions where CKE is reduced to create MKE:  (1) immediately behind the cylinder and around $x/>10$ along the centerline. However, the total production integrated over the domain is positive indicating that the energy cascade to the turbulent scales from the mean flow exists.  Figure \ref{fig:production}(b) shows the random production rate $\mathcal{P}_r$, which captures the production of RKE from the mean scales and appears in the MKE and RKE equations. Overall, the random production occurs in the wake and is predominately positive indicating energy transfers from the mean flow to the random turbulent scales.   The scale of the production is two orders of magnitude smaller that the coherent production further indicating that the effects of the RKE are small compared the CKE.  There is a third production term, which accounts for the transfer of coherent energy to random energy and is shown in Fig. \ref{fig:production}(c). This term appears in both the CKE and RKE equations.  While relatively small compared to the coherent production, there exists an exchange of energy in the wake. In the near wake, where the turbulence is highest, the production is positive indicating transfer from CKE to RKE.  However, further downstream, the energy moves from CKE to RKE as the formation of von K\'{a}rm\'{a}n vortices become strong and coherent. 

%
%
\bibliographystyle{jfm}
\bibliography{./bibl}
\end{document}